\documentclass[a4paper]{aa}

\usepackage{graphicx,natbib}
\usepackage{txfonts}
\usepackage{color}
\usepackage[usenames,dvipsnames]{xcolor}

\newcommand{\bz}{$\langle B_\mathrm{z} \rangle$}
\newcommand{\kms}{km\,s$^{-1}$}
\newcommand{\ms}{m\,s$^{-1}$}
\newcommand{\reduce}{{\sc reduce}}
\newcommand{\cla}[1]{#1}

\graphicspath{{fig/}}

\begin{document}
\title{\cla{Three-dimensional magnetic and abundance mapping \\ of the cool Ap star HD\,24712}}

\subtitle{\cla{I. Spectropolarimetric observations in all four Stokes parameters%
\thanks{Based on observations collected at the European Southern Observatory, Chile (ESO programs 084.D-0338, 085.D-0296, 086.D-0240).}}}

\author{N.~Rusomarov\inst{1}
   \and O.~Kochukhov\inst{1}
   \and N.~Piskunov\inst{1}
   \and S.~V.~Jeffers\inst{2}
   \and C.~M.~Johns-Krull\inst{3}
   \and C.~U.~Keller\inst{2}
   \and V.~Makaganiuk\inst{1}
   \and M.~Rodenhuis\inst{2}
   \and F.~Snik\inst{2}
   \and H.~C.~Stempels\inst{1}
   \and J.~A.~Valenti\inst{5}}

\institute{
Department of Physics and Astronomy, Uppsala University, Box 516, 75120 Uppsala, Sweden
\and
Institute of Astrophysics, Georg-August University, Friedrich-Hund-Platz 1, D-37077 G\"ottingen, Germany
\and
Department of Physics and Astronomy, Rice University, 6100 Main Street, Houston, TX 77005, USA
\and
Sterrewacht Leiden, Universiteit Leiden, Niels Bohrweg 2, 2333 CA Leiden, The Netherlands
\and
Space Telescope Science Institute, 3700 San Martin Dr, Baltimore MD 21211, USA
}

\date{Received 00 December 2012 / Accepted 00 December 2012}

\titlerunning{Spectropolarimetry of HD\,24712 in all four Stokes parameters}
\authorrunning{N.~Rusomarov et al.}

  \abstract
{High-resolution spectropolarimetric observations provide simultaneous information about stellar magnetic field topologies and three-dimensional distributions of chemical elements. High-quality spectra in the Stokes $IQUV$ parameters are currently available for very few early-type magnetic chemically peculiar stars. Here we present analysis of a unique full Stokes vector spectropolarimetric data set, acquired for the cool magnetic Ap star HD\,24712 with a recently commissioned spectropolarimeter.}
{The goal of our work is to examine the circular and linear polarization signatures inside spectral lines and to study variation of the stellar spectrum and magnetic observables as a function of rotational phase.}
{HD\,24712 was observed with the HARPSpol instrument at the 3.6-m ESO telescope over a period of 2010--2011. We achieved full rotational phase coverage with 43 individual Stokes parameter observations. The resulting spectra have S/N ratio of 300--600 and resolving power exceeding $10^5$. The multiline technique of least-squares deconvolution (LSD) was applied to combine information from the spectral lines of Fe-peak and rare-earth elements.}
{We used the HARPSPol spectra of HD\,24712 to study the morphology of the Stokes profile shapes in individual spectral lines and in LSD Stokes profiles corresponding to different line masks. From the LSD Stokes~$V$ profiles we measured the longitudinal component of the magnetic field, \bz, with an accuracy of 5--10\,G. We also determined the net linear polarization from the LSD Stokes $Q$ and $U$ profiles. Combining previous \bz\ measurements with our data allowed us to determine an improved rotational period of the star,  $P_\mathrm{rot} = 12.45812 \pm 0.00019$~d. We also measured the longitudinal magnetic field from the cores of H$\alpha$ and H$\beta$ lines. The analysis of \bz\ measurements showed no evidence for a significant radial magnetic field gradient in the atmosphere of HD\,24712. We used our \bz\ and net linear polarization measurements to determine parameters of the dipolar magnetic field topology. We found that magnetic observables can be reasonably well reproduced by the dipolar model, although significant discrepancies remain at certain rotational phases. We discovered rotational modulation of the H$\alpha$ core and related it a non-uniform surface distribution of rare-earth elements.}
{}


\maketitle

\section{Introduction}
\label{sec:intro}
Chemically peculiar (CP) stars comprise less than 20\% of the intermediate and upper main sequence stars \citep{Catalano94} and are characterized by rich spectra. Relative to the solar abundance these stars often show overabundances of up to few dex for some iron peak elements and rare earth elements, while other chemical elements are found to be underabundant \citep{Ryabchikova2004}. A subset of these stars are the magnetic Ap stars. These stars show enhanced abundances of elements such as Si, Cr, Sr and Eu, and generally are slow rotators. The magnetic fields that these stars exhibit have strengths from a few hundred G up to tens of kG, are static on timescales of many decades, and appear to be frozen into a rigidly rotating stellar atmosphere. Many spectral lines in these stars show prominent and strictly periodic changes that can be explained by the oblique magnetic rotator model \citep{Stibbs1950p395}, which attributes the observed variability to a modulation of the aspect angle at which an inhomogeneous surface structure is seen by the observer. The magnetic fields present in these stars have significant influence on the energy and mass transport within the atmosphere, and are also responsible for the creation of abundance patches and vertical chemical stratification \citep{Khan2006,Alecian2010}. Atmospheric processes, such as atomic diffusion \citep{Michaud1970p641}, magnetically driven stellar winds \citep{Babel1997} and convection, should play important roles but are still poorly understood.

An important subset of Ap stars are the rapidly oscillating Ap stars (roAp) which pulsate with periods of 6--24 minutes \citep{Kurtz2000p253,Alentiev2012}. Their observed pulsation amplitudes and phases are modulated according to the magnetic field variations \citep[e.g.][]{Kurtz1982p807,Kochukhov2006}. These stars experience pulsations in non-radial p-modes of very high overtone, with their pulsation and magnetic axes aligned. A complex interplay between the abundance patches, magnetic field and pulsations is still little investigated but, in principle, can be readily studied for roAp stars because their magnetic and pulsational characteristics can be very well constrained observationally. Therefore these objects are ideal laboratories for asteroseismology and for detailed analysis of the structure of magnetic field and its influence on the atomic diffusion and pulsations.

The most powerful methods used in previous investigations of roAp-star atmospheres include magnetic Doppler imaging \citep{Piskunov2002p736}, spectroscopic and photometric pulsational studies \citep{Ryabchikova2007p1103} and self-consistent empirical stratified model atmosphere analysis \citep{Shulyak09p879}.

The goal of magnetic Doppler imaging \citep[MDI,][]{Piskunov2002p736,Kochukhov2002p868} is to reconstruct horizontal distributions of different chemical elements simultaneously with obtaining a vector map of the magnetic field on the stellar surface. MDI was applied to investigate surface inhomogeneities and magnetic fields in Ap/Bp stars \citep[e.g.][]{Kochukhov10p13,Luftinger10p71} and in cool active stars \citep[e.g.][]{Donati2003}. Most previous MDI studies of Ap stars relied on only the circular polarization data for magnetic mapping, requiring multipolar regularization or parameterization of the magnetic field topology. With the advent of high-resolution spectropolarimetry in all four Stokes parameters this limitation of the MDI method was lifted, thus enabling a retrieval of self-consistent, assumption-free maps of the vector magnetic fields and chemical abundance spots. These improvements in observational data have led to a number of interesting new results, including the detection of small-scale surface magnetic structures previously not known for Ap stars \citep{Kochukhov2004,Kochukhov10p13}.

Spectroscopic and photometric pulsational analysis is another prominent method, well-suited for probing the interior of stars, that is capable of testing theories of stellar evolution and structure. Spectroscopic and photometric techniques are able to provide information about the different pulsation modes and hence atmospheric layers at different optical depths in the stellar atmosphere. Some investigators \citep[e.g.][]{Ryabchikova2007p1103} have used radial velocity amplitudes of different spectral lines, measured from spectroscopic data, as a function of optical depth to describe the properties of pulsational waves in the atmospheric layers where these spectral lines are formed. Although this type of analysis is only possible for a few bright roAp stars, such investigations have demonstrated that for understanding the phenomenon of roAp stars a good knowledge of the vertical structure of their surface layers is essential.

A complementary observational approach to investigating the vertical segregation of chemical elements is iterative fitting of observed high-resolution spectra and energy distribution using model atmospheres that properly account for the highly abnormal chemical properties of Ap stars \citep[see e.g.][]{Shulyak09p879,Kochukhov2009}. These investigations have become possible with the recent development of a stellar model atmosphere code that can produce 1-D, magneto-hydrostatic, LTE atmospheres with individual and vertically stratified abundances \citep{Shulyak2004p993}.

The three methods discussed above have been instrumental in deepening our understanding of the atmospheres of Ap stars. The studies done with their help have unveiled a new level of complexity of stellar atmospheres, at the same time providing essential observational constraints for the theories of radiative diffusion in a strong magnetic field. However, the general coherent picture is still lacking since previous studies typically addressed different aspects of the chemical and magnetic structure of different Ap stars. Furthermore, the current methods that are available at our disposal suffer from various limitations. Current MDI codes, although very powerful, do not incorporate the vertical chemical stratification. On the other hand, the empirical stratified model atmospheres use only intensity spectra and do not consider horizontal variation due to stellar surface structure. 

We believe that these limitations of the current methods can be overcome by a simultaneous study of 3-D magnetic and chemical structures using the best currently available four Stokes parameter spectropolarimetric observations. For that purpose we have started a program aimed at observing Ap stars in all four Stokes parameters with the newly developed HARPSpol polarimeter. As the first target we have chosen the star HD\,24712, which is the best studied roAp star and one of the coolest known Ap stars showing both chemical stratification signatures and oblique rotator variations. This paper is the first in a series of publications where we will explore the possibility of performing 3-D MDI analysis. Here we focus on evaluation of the new four Stokes parameter spectropolarimetric observations and  analysis of circular and linear polarization in spectral lines. Subsequent papers will present magnetic and chemical inversions using these data.

Our paper is composed in the following way. In Sect.~\ref{sec:hd24712} we summarize the main properties and previous studies of our target star, HD\,24712. In Sect.~\ref{sec:obs} we describe  spectropolarimetric observations and data reduction. Sect.~\ref{sec:ilp} and \ref{sec:lsd} discuss polarization signatures in individual spectral lines and in mean profiles respectively. We present longitudinal magnetic field measurements in Sect.~\ref{sec:mag}. Section~\ref{sec:rot} is devoted to the search rotational period of HD\,24712. The paper ends with Sect.~\ref{sec:sum} where we present the summary of key results and discuss a possibility of magnetic field gradient in the atmosphere of HD\,24712.

\section{HD\,24712}
\label{sec:hd24712}

HD\,24712 (HR\,1217, DO Eri, HIP\,18339) first appears in the catalog of A stars with peculiar lines by \citet{Bertaud1959p45}. In the paper by \citet{Preston1970p878} HD\,24712 is listed together with Sr-Cr-Eu stars that are known to have $v_{\rm e}\sin i \le 10$\,\kms{}. Magnetic field measurements by \citet{Preston1972p465} showed that the mean longitudinal magnetic field of HD\,24712 changes with a period of $\approx$\,12.45 days. This variability is accompanied by spectral line changes, for example an antiphase variations between the lines of Mg and Eu. The variability period found by \citet{Preston1972p465} was confirmed by the photometric observations by \citet{Wolff1973p141}. Many studies of the magnetic field of HD\,24712 have been carried out \citep{Preston1972p465,Brown1981p899,Mathys1991p121,Mathys1994p547,Mathys97p475,Ryabchikova97p1137,Leone2000p315,Wade00p851,Leone2004p271,Ryabchikova2005p55,Ryabchikova2007p1103}. Below we will use some of these measurements for improving rotational period of the star.

The discovery of rapid oscillations \citep{Kurtz1982p807} with multiple periods near 6.13 min made HD\,24712 one of the prototype rapidly oscillating Ap (roAp) star and arguably the most asteroseismically interesting object in this class. \citet{Matthews1988p1099} detected radial velocity oscillations with an amplitude of $400 \pm 50\,\mathrm{m/s}$ and with a period close to the photometric one. More recently, spectroscopic studies of pulsations in HD\,24712 were carried out by \citet{Mkrtichian2005}, \citet{Ryabchikova2007p1103} and \citet{Kochukhov2007a}. Time-resolved magnetic field measurements by \citet{Kochukhov2007b} did not reveal any variability of the magnetic field with pulsational phase in this star. HD\,24712 was targeted by the Whole Earth Telescope (WET) campaign \citep{Kurtz2002p57} and by the high-precision space photometry using MOST satellite \citep{Ryabchikova2007p1103}. The wealth of asteroseismic information provided by the multi-periodic pulsations in this star has led to significant advances in understanding of the impact of magnetic field on the excitations and geometry of non-radial oscillations in in the presence of strong magnetic field \citep[e.g.][]{Cunha2003p831,Saio2010}.

Magnetic Doppler Imaging (MDI) of HD\,24712 was performed by \citet{Luftinger10p71} using Stokes $I$ and $V$ spectra. In this study the magnetic field geometry was determined from seven \ion{Fe}{i} and five \ion{Nd}{iii} lines which have strong polarization signatures and are not affected by blending. From these data a surface magnetic field with dipole structure was derived, with the local magnetic field strength varying between 2.1\,kG and 4.2\,kG. These results agree well with the dipolar model derived by \citet{Bagnulo1995p459} from the phase curves of the longitudinal field and broad-band linear polarization. Abundance distributions of Fe, Nd and sixteen additional ions were also determined by \citet{Luftinger10p71}. They found that Fe is slightly underabundant compared to the solar value ($\log N_{\rm Fe}/N_{\rm tot}$\,=\,$-4.59$), with local abundances varying between $-5.3$ and $-4.7$. Element concentration derived from \ion{Nd}{iii} lines is strongly overabundant relative to the solar value ($\log N_{\rm Nd}/N_{\rm tot}$\,=\,$-10.59$) and does not match the abundance derived from \ion{Nd}{ii} which is 1.5\,dex smaller. A spatial anticorrelation between the distributions of Fe and Nd was also found, as expected from the equivalent width variations of these elements \citep{Bonsack1979p648}. In HD\,24712 Fe is overabundant where one finds a relative depletion of Nd and vice versa. REEs accumulate in the region close to the positive magnetic pole, while the Fe-peak elements accumulate in the magnetic equatorial region. At the same time, non-negligible differences between surface distributions of similar elements were also observed.

High-resolution spectroscopic radial velocity measurements by \citet{Ryabchikova2007p1103} probed different line formation depths in the atmosphere of HD\,24712 and allowed to trace propagation of the pulsational waves in the outer atmosphere of this star. It was found that REE lines and H$\alpha$ core have large pulsation amplitudes (150--400\,\ms{}), while Mg, Si, Ca and Fe-peak elements show no detectable variability. This diversity of pulsational behaviour points at strong vertical inhomogeneities in stellar atmosphere. The conclusions from the pulsation studies were directly supported by the analysis based on empirical, chemically stratified model atmosphere calculations performed by \citet{Shulyak09p879}. This study showed that, similar to many other cool Ap stars \citep[e.g.][]{Ryabchikova2002}, the majority of elements show nearly solar or underabundant concentrations in the upper layers but exhibit a higher abundance in the layers around the photosphere of the star. In contrast, the vertical abundance distribution of Pr and Nd shows a large overabundance (up to 5\,dex) above $\log\tau_{5000}\approx-3$ and approximately solar abundance in the photospheric layers. It is worth noting that REE stratification can significantly influence the model atmosphere and even lead to appearance of an inverse temperature gradient in the upper atmospheric regions. The presence of this temperature anomaly may influence the formation of the inner wings and the core of the hydrogen H$\alpha$ line.

The basic physical parameters of HD\,24712, compiled from the literature, are listed in Table~\ref{tab:stellar-parameters}.
\begin{table}
  \caption{Basic parameters of HD\,24712}
  \centering
  \begin{tabular}{llc}
    \hline
    \hline
Parameter & Value & Reference \\
    \hline
    $T_\mathrm{eff}$ & $7250 \pm 150\,\mathrm{K}$ & 1, 2 \\
    $\log g$ & $4.2 \pm 0.1$ & 1, 2 \\
    $R/R_\odot$ & $1.772 \pm 0.043$ & 2 \\
    $\log (L/L_\odot)$ & $0.891 \pm 0.041$ & 3 \\
    $M/M_\odot$ & $1.55 \pm 0.03$ &  4 \\ 
    $P_\mathrm{puls}$ & $\approx 6.13\,\mathrm{min}$ & 5, 6 \\
    $v_{\rm e}\sin i$ & $5.6\pm2.3$\,\kms & 1 \\
     $P_\mathrm{rot}$ & $12.45877 (16)\,\mathrm{days}$ & 7 \\
    \hline
  \end{tabular}
  \label{tab:stellar-parameters}
\tablebib{
(1) \citet{Ryabchikova97p1137}; (2) \citet{Shulyak09p879}; (3) \citet{Matthews1999p422}; (4) \citet{Kochukhov2006p763}; (5) \citet{Kurtz2002p57}; (6) \citet{Ryabchikova2007p1103}; (7) \citet{Ryabchikova2005p55}; 
}  
\end{table}

\section{Observations and data reduction}
\label{sec:obs}

\begin{table*}
  \caption{Journal of spectropolarimetric observations of HD\,24712.}
  \centering
  \begin{tabular}{lcccccccccccc}
    \hline
    \hline
    UT Date & \multicolumn{3}{c}{Stokes} & \multicolumn{3}{c}{HJD\,(2\,455\,000+)} & & & S/N & \multicolumn{3}{c}{median S/N}\\
    & \multicolumn{3}{c}{Parameters} & $Q$ & $U$ & $V$ & $\overline{\varphi}$ & $\delta\varphi$ & range & $Q$ & $U$ & $V$\\
    \hline
    4 Jan. 2010 & $Q$ & $U$ & $V$ & 200.7184 & 200.6962 & 200.7421 & 0.772 & 0.002 & 121--384 & 309 & 343 & 224\\
    5 Jan. 2010 & $Q$ & $U$ & $V$ & 201.7130 & 201.7364 & 201.7561 & 0.853 & 0.002 &  95--342 & 287 & 300 & 181\\
    6 Jan. 2010 & $Q$ & $U$ & $V$ & 202.6599 & 202.6240 & 202.6981 & 0.928 & 0.004 & 191--458 & 405 & 326 & 387\\
    7 Jan. 2010 & $Q$ & $U$ & $V$ & 203.6702 & 203.7121 & 203.6274 & 0.009 & 0.004 & 235--567 & 504 & 408 & 484\\
    8 Jan. 2010 & $Q$ & $U$ & $V$ & 204.6082 & 204.6546 & 204.7039 & 0.088 & 0.005 & 303--586 & 528 & 529 & 497\\
    9 Jan. 2010 & $Q$ & $U$ & $V$ & 205.6338 & 205.6802 & 205.7272 & 0.170 & 0.005 & 170--483 & 429 & 379 & 307\\
   10 Jan. 2010 & $Q$ & $U$ & $V$ & 206.6092 & 206.6560 & 206.7032 & 0.248 & 0.005 & 267--565 & 498 & 506 & 457\\
   11 Jan. 2010 & $Q$ & $U$ &     & 207.6547 & 207.7014 &          & 0.330 & 0.003 & 261--510 & 440 & 451 &    \\
   13 Jan. 2010 & $Q$ & $U$ & $V$ & 209.6057 & 209.6518 & 209.6990 & 0.489 & 0.005 & 181--571 & 512 & 480 & 382\\
   14 Jan. 2010 & $Q$ & $U$ & $V$ & 210.6245 & 210.6697 & 210.7157 & 0.570 & 0.005 & 200--573 & 510 & 461 & 360\\
   15 Jan. 2010 & $Q$ & $U$ & $V$ & 211.6183 & 211.6683 & 211.7185 & 0.651 & 0.005 & 201--528 & 454 & 469 & 365\\
   16 Jan. 2010 & $Q$ & $U$ & $V$ & 212.6248 & 212.6704 & 212.7052 & 0.731 & 0.004 & 189--580 & 512 & 517 & 330\\
   17 Jan. 2010 & $Q$ & $U$ & $V$ & 213.6272 & 213.6723 & 213.7132 & 0.811 & 0.004 & 162--607 & 545 & 458 & 297\\
   13 Aug. 2010 &     &     & $V$ &          &          & 421.8958 & 0.525 & 0.001 & 168--336 &     &     & 302\\
   14 Feb. 2011 &     &     & $V$ &          &          & 606.5350 & 0.346 & 0.001 & 278--497 &     &     & 446\\
   15 Feb. 2011 & $Q$ & $U$ & $V$ & 607.5200 & 607.5541 & 607.5882 & 0.428 & 0.003 & 223--508 & 451 & 417 & 394\\
   \hline
  \end{tabular}
  \label{tab:obs-journal}
  \tablefoot{First column gives the UT date at the beginning of each observing night. Column 2 indicates the Stokes parameters observed (in addition to Stokes~$I$). Heliocentric Julian Dates (HJD) at mid-exposure for each observed Stokes parameter are given in columns 3--5. Mean phase, $\overline{\varphi}$, and the maximum difference, $\delta\varphi$, between $\overline{\varphi}$ and phases of individual Stokes parameter observations are presented in columns 6--7. Rotational phases were calculated according to our improved ephemeris (Sect.~\ref{sec:rot}). The range of the signal to noise (S/N) ratio for two or three Stokes parameter observations and the median S/N ratio for each individual Stokes parameter observation are presented in columns 8--11. The median S/N ratio was calculated using several orders around $\lambda=5500$\,\AA, where maximum counts were reached.}
\end{table*}

We obtained the spectra of HD\,24712 with the HARPSpol polarimeter \citep{Snik2011p237, Piskunov11p7} that feeds the HARPS spectrometer \citep{Mayor2003p20} at the ESO 3.6-m telescope at La Silla, Chile. The two optical fibres transport the light, split into two orthogonal polarization states, from the polarimeter at the Cassegrain focus of the telescope to the HARPS spectrograph. The HARPSpol instrument consists of two independent polarimeters for measurements in circular and linear polarization.  Together, the two polarimeters allow observations in all four Stokes parameters. Each polarimeter consists of a polarizing beam splitter, in the form of a Foster prism fixed relative to the fibres and a super-achromatic rotating retarder wave plate placed in front of it. The circular polarimeter has a quarter-wave plate and the linear polarimeter is equipped with a half-wave plate. The wave plates used in HARPSpol are zero-order birefringent polymer retarders, which are not affected by the interference fringes typical of the crystalline super-achromatic wave plates \citep{Samoylov2004}. More information on the optical design of HARPSpol can be found in the paper by \citet{Snik2008p}.

A spectropolarimetric observation of one Stokes parameter consists of four sub-exposures taken at different orientations of the half-wave or quarter-wave plate. Each sub-exposure corresponds to a different position of the wave plate relative to the optical axis of the beam splitter. For Stokes~$Q$ the exposures are taken at positions of the half-wave plate: $0^\circ$, $45^\circ$, $90^\circ$, $135^\circ$. For  Stokes~$U$ the positions of the half-wave plate are $22.5^\circ$, $67.5^\circ$, $112.5^\circ$, $157.5^\circ$, and finally, for Stokes~$V$ the positions of the quarter-wave plate are $45^\circ$, $135^\circ$, $225^\circ$ and $315^\circ$. A given Stokes parameter can be also obtained with only two exposures at two different orientations of the wave plate, however this does not allow deriving a diagnostic null spectrum and assessing spurious polarization and crosstalks. Therefore, most individual Stokes parameter observations of HD\,24712 were obtained with four sub-exposures. A small number of observations were obtained using two exposures, mostly due to time constraints. 

The star was observed over the period of 2010--2011, resulting in an excellent rotational phase coverage. In total we have obtained 43 individual Stokes parameter observations during three observing runs spread over 16 nights. For thirteen nights we have Stokes~$IQUV$ observations, while for two nights we have Stokes~$IV$ and for one night we have Stokes~$IQU$ observations. The majority of the Stokes parameter observations were obtained in January 2010 (38 in total), while the rest was obtained in August 2010 and February 2011. 

The HARPSpol spectra have the resolving power of $\lambda/\Delta\lambda\approx110\,000$ and cover the wavelength range 3800--6910\,\AA{} with a 80\,\AA{} gap located at 5290\,\AA. Each Stokes parameter observation was calculated from two or four sub-exposures, each with individual exposure time between 440 and 1040 seconds and the median exposure time around 940 seconds. During each observing night two sets of 10 flat frames, 10 bias exposures and one ThAr image were acquired before the start and at the end of the observations. Table~\ref{tab:obs-journal} provides detailed summary of our spectropolarimetric observations of HD\,24712. 

For the reduction of the spectra we used the \reduce{} package \citep{Piskunov02p1095}. This software package is written in the Interactive Data Language (IDL) and is designed for processing cross-dispersed echelle spectra. The reduction process consists of a number of standard steps. At the start of the procedure, the mean bias and flat frames are constructed, and the bias frame is subtracted from the latter. The mean flat frame is used for construction of the normalized flat field. The position and shape of the spectral orders is determined from the mean flat field with the help of a cluster analysis. The curvature of the spectral orders is approximated by a 4th-order polynomial function.
The science frames are initially processed by subtracting the mean bias frame and dividing by the normalized flat field, which corrects the small-scale pixel-to-pixel sensitivity variations. After the scattered light is estimated and subtracted from the science frames, an optimal extraction algorithm \citep{Piskunov02p1095} is applied for the extraction of 1-D spectra.

The exceptional velocity stability of the HARPS spectrograph, at the level of 1\,\ms{} as reported by \citet{Mayor2003p20}, and the nature of our scientific goals allow using a single ThAr spectrum for the wavelength calibration procedure during each night. By measuring around 3000 ThAr lines, we obtained 2-D wavelength solutions separately for the blue and red HARPS CCD detectors with an internal accuracy of 15--20\,\ms. The last step of the reduction procedure involves continuum normalization. For this purpose we first corrected the extracted spectra by the blaze function inferred from the mean flat field and then divided the spectrum by a smooth slowly varying function, that was obtained by fitting the upper envelope of the spectrum.

The Stokes parameters were derived from four sub-exposures with the ``ratio method'' described by \citet{Donati97p658} and \citet{Bagnulo2009p993}. This polarimetric demodulation technique is able to remove spurious polarization effects up to first order by combining the two ordinary and extraordinary beams recorded at four different positions of the wave plate. The beams can be also combined in such a way that intrinsic polarization cancels out. The resulting diagnostic null spectrum is a valuable quality control tool since it allows assessing crosstalks between Stokes parameters and diagnosing other sources of spurious polarization. In a couple of cases when only two sub-exposures were available a truncated version of the ratio method was used that did not provide the null spectrum.
\cla{A detailed assessment of the spurious polarization and cross-talk between different Stokes parameters is presented in Appendix~\ref{sec:crosstalk}.}

\section{Stokes profiles of individual spectral lines}
\label{sec:ilp}

\begin{figure*}[!t]
  \centering
  \includegraphics[width=\textwidth]{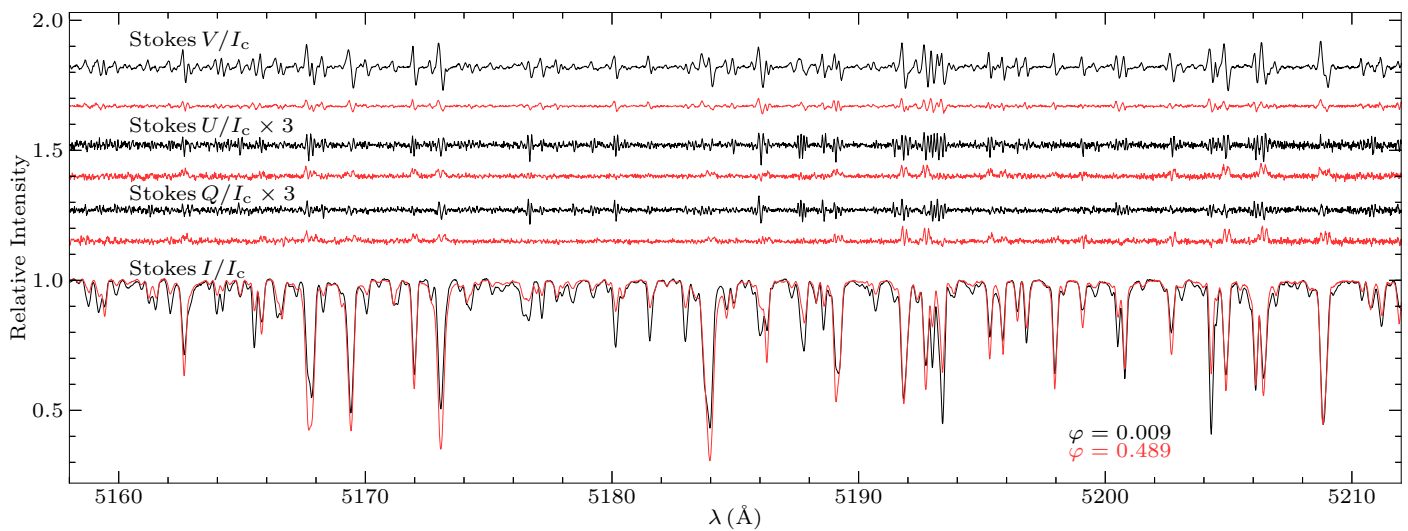}
  \caption{HARPSpol four Stokes parameter spectra of HD\,24712 in the 5190--5210\,\AA{} wavelength region. The spectra plotted with black lines were obtained around zero phase, when the longitudinal magnetic field reaches its maximum. The spectra plotted with red lines were obtained close to phase 0.5 corresponding to the magnetic minimum. The circular and linear polarization profiles are offset vertically for clarity. Note that Stokes $Q$ and $U$ spectra have been expanded by a factor of 3 compared to Stokes $V$ and $I$.
}
  \label{fig:StokesParameters}
\end{figure*}

Previous systematic, phase-resolved observations of Ap stars in all four Stokes parameters were carried out at the resolving power of only $\approx$\,35\,000 \citep{Wade00p823}. Recently, this work was extended to $R=65\,000$ \citep{Silvester2012}. Here we present \textit{the first ever full Stokes vector data set at $R>10^5$}. These data provide us with a unique opportunity to measure and model high-quality Stokes profiles of individual spectral lines across a wide wavelength range. The Ap-star target chosen for our analysis turns out to be especially suitable for the spectropolarimetric monitoring at a very high spectral resolution due to its moderately strong magnetic field, rich metal line spectrum and low $v_{\rm e}\sin i$. We found that the amplitude of linear polarization signatures in this star increases by a factor of two by going from $R=65\,000$ (ESPaDOnS, Narval) to 110\,000 (HARPSpol).

Thanks to a combination of high resolution and high S/N, we are able to detect linear polarization signatures in numerous intermediate-strength metal lines, whereas in previous studies of Ap stars this was possible to achieve only for a handful of strongest lines \citep{Wade00p823}. In Fig.~\ref{fig:StokesParameters} we present the observed spectra of HD\,24712 in all four Stokes parameters in the 5160--5210\,\AA{} region. This figure compares two four Stokes profile observations obtained at magnetic maximum and minimum, respectively. One can see that many individual spectral lines show strong and complex linear polarization signatures, especially at magnetic maximum. In general, we were able to identify several dozen unblended spectral lines which exhibit clear $Q$ and $U$ signatures at all rotational phases. Many more lines showing significant linear polarization signatures are parts of complex blends. A few examples of rotational modulation of the four Stokes parameter profiles of individual lines can be found in Fig.~\ref{fig:StokesProfilesNdFe} (isolated \ion{Fe}{i} and \ion{Nd}{iii} lines) \cla{and Fig.~\ref{fig:StokesProfiles-5190-5196} in the online material (several Fe-peak and REE lines)}. Note the Stokes $I$ profile variations of the magnetically sensitive ($\overline{g}=2.01$) \ion{Fe}{i} 4938.81~\AA\ line in the latter plot. This line is clearly broader and more distorted due to a stronger Zeeman splitting at phase 0.0 corresponding to the magnetic field maximum.

From the visual analysis of unblended spectral lines we find that spectral lines of the Fe-peak elements tend to show a weaker signature in Stokes~$Q$, compared to the lines of REEs. The shape of the Stokes profiles of Fe-peak element and REE lines also looks different and exhibits a different rotational modulation. This may be attributed to different element distributions across the surface of the star. Variation of the vertical abundance profiles and possible vertical magnetic field gradient may also contribute to the observed behaviour of Stokes parameters of different species. Interestingly, while the amplitude of these signatures varies strongly with phase, the shape of the Stokes~$Q$ and $U$ profiles changes less dramatically as a function of phase in individual lines. 

Stokes~$I$ appears to vary for many spectral lines as a function of phase. This is evident from Fig.~\ref{fig:StokesParameters}, where one can see significant differences for many lines between the Stokes~$I$ spectra corresponding to magnetic minimum (strong Fe-peak element lines) and maximum (strong REE lines). In general, it appears that most spectral lines exhibit strong intensity variation. The most extreme example is provided by the \ion{Na}{i} D-lines at 5890 and 5896~\AA. While inconspicuous in many Ap stars, this element apparently exhibits an extreme horizontal gradient in HD\,24712, leading to the equivalent width variation from $<30$~m\AA\ at phase 0 to $\approx$\,150~m\AA\ at phase 0.5 for the D1 line.

\section{Least-Squares Deconvolution}
\label{sec:lsd}

\begin{figure*}
  \centering
  \includegraphics[width=0.4975\textwidth]{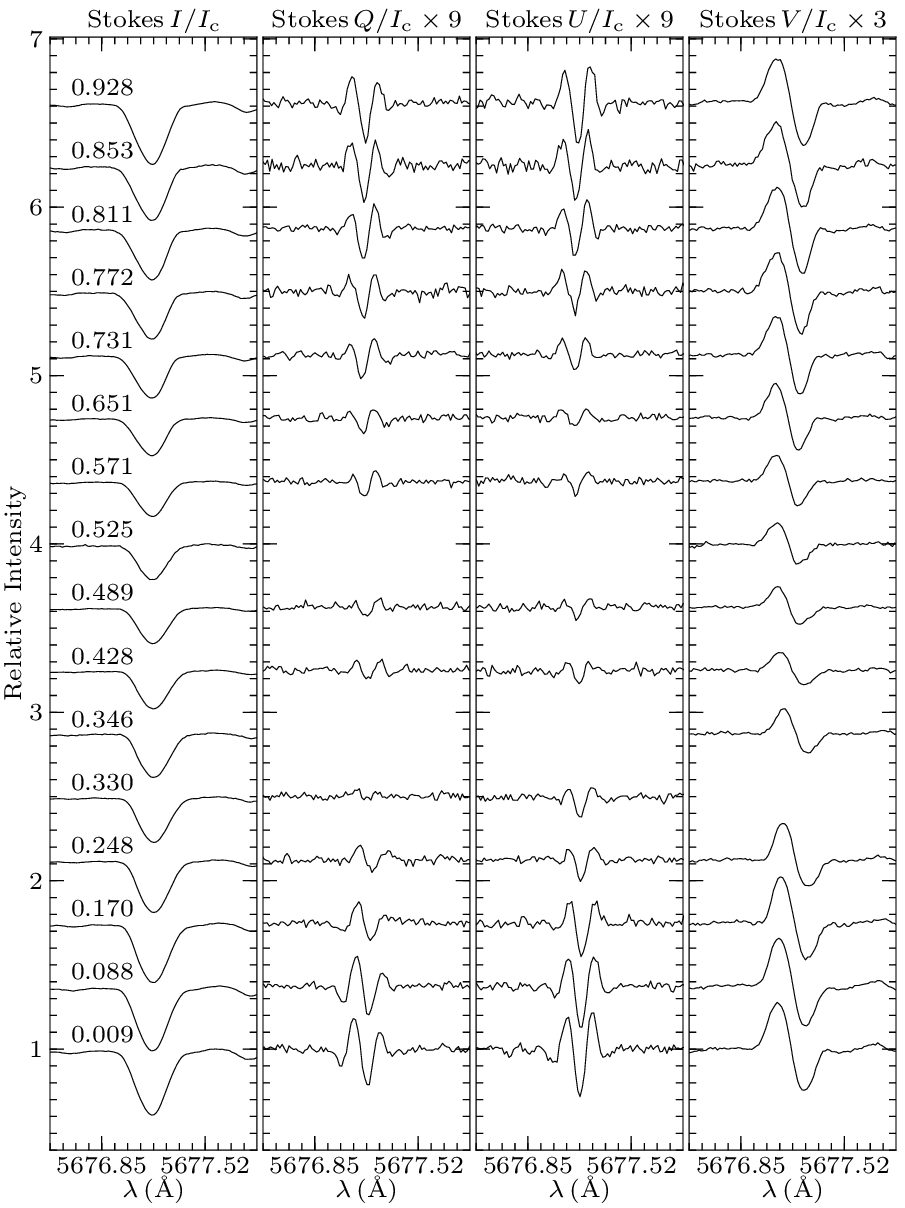}
  \includegraphics[width=0.4975\textwidth]{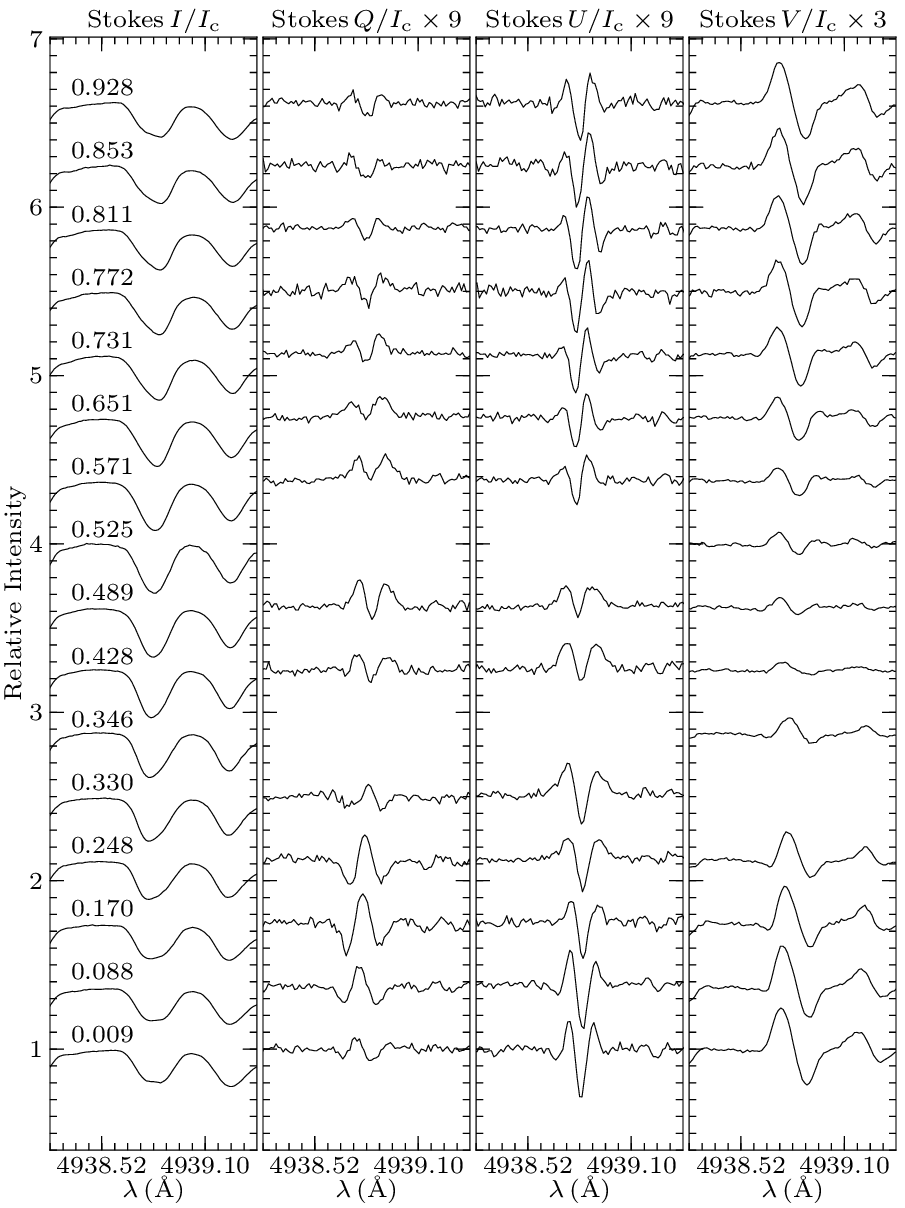}
  \caption{Rotational modulation of the Stokes~$IQUV$ profiles for the \ion{Nd}{iii} 5677.18~\AA{} (left panel) and \ion{Fe}{i} 4938.81~\AA{} (right panel) spectral lines. \cla{The spectra for different rotation phases are offset vertically. The Stokes $QU$ and $V$ profiles are expanded by factors 9 and 3, respectively, relative to Stokes $I$. Rotational phases are indicated in the left panel.}
  }
  \label{fig:StokesProfilesNdFe}
\end{figure*}

\onlfig{
\begin{figure*}
  \centering
  \includegraphics{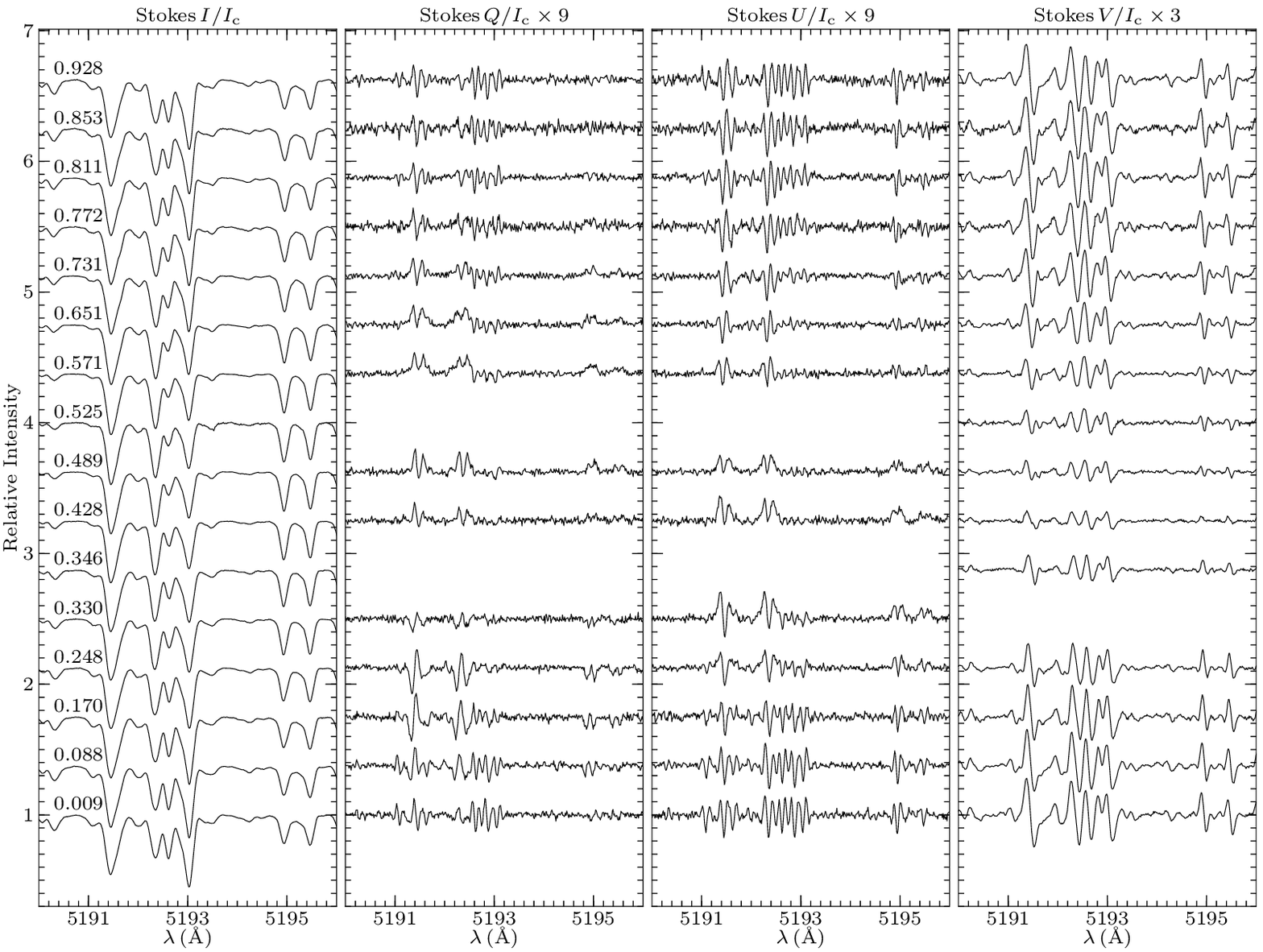}
  \caption{Variations of Stokes~$IQUV$ profiles in the 5190--5196\,\AA{} wavelength region. The four prominent features at 5191--5193\,\AA{} are blends of Fe-peak and REE lines. The other two lines around 5195\,\AA{} belong to Fe. This plot is characteristic of the complex polarization profiles observed in the spectra of HD\,24712.  \cla{The format of this figure is similar to Fig.~\ref{fig:StokesProfilesNdFe}.}}
  \label{fig:StokesProfiles-5190-5196}
\end{figure*}
}

Although the quality of the HARPSpol Stokes spectra of HD\,24712 is very high, allowing us to analyse polarization profiles of individual spectral lines, it is still useful to employ multiline diagnostic methods in order to deduce mean characteristics of the stellar magnetic field and be able to compare our results with previous studies of this star. To this end, we used the technique of Least-Squares Deconvolution (LSD, see \citealt{Donati97p658} and references therein). \citet{Wade00p851} demonstrated the usefulness of the LSD approach to measurements of the mean longitudinal magnetic field and  net linear polarization in Ap stars. LSD assumes that the stellar intensity or polarization spectrum can be represented as a superposition of scaled, self-similar profiles and that overlapping lines add up linearly. The robustness and limitations of the LSD technique have been investigated by \citet{Kochukhov10p5}.

For the calculation of the mean profiles of HD\,24712 we employed the multiprofile LSD code and methodology described by \citet{Kochukhov10p5}. Multiprofile LSD is capable of disentangling the mean profiles of specific elements, while minimizing the effects of blends from the lines of other elements. This technique is of fundamental importance for our study as magnetic chemically peculiar stars exhibit diverse distributions of chemical elements, which results in different mean profiles for different species of chemical elements.

The line mask for the LSD procedure was extracted from the VALD database \citep{Piskunov1995p525,Kupka1999p119} for the stellar parameters presented in Table~\ref{tab:stellar-parameters}. The abundances used for the calculation of the line mask are mean values of the abundances taken from \citet{Luftinger10p71} and \citet{Ryabchikova97p1137}. From the initial line mask containing 10992 lines we removed the spectral lines affected by the hydrogen and telluric lines, and the lines with central depth less than 10\% of the continuum, thus yielding a final mask with 5120 lines. This line list is dominated by \ion{Fe}{i} (876 lines), with an average wavelength $\lambda=4973$\,\AA{}, mean Land\'e factor $\overline{g}=1.292$ and mean residual intensity $d=0.349$. Among rare-earth elements (REEs) the most common is \ion{Ce}{ii} (803 lines), with $\lambda=4666$\,\AA{}, $\overline{g}=1.072$ and $d=0.285$.

\begin{figure*}[!t]
  \centering
  \includegraphics{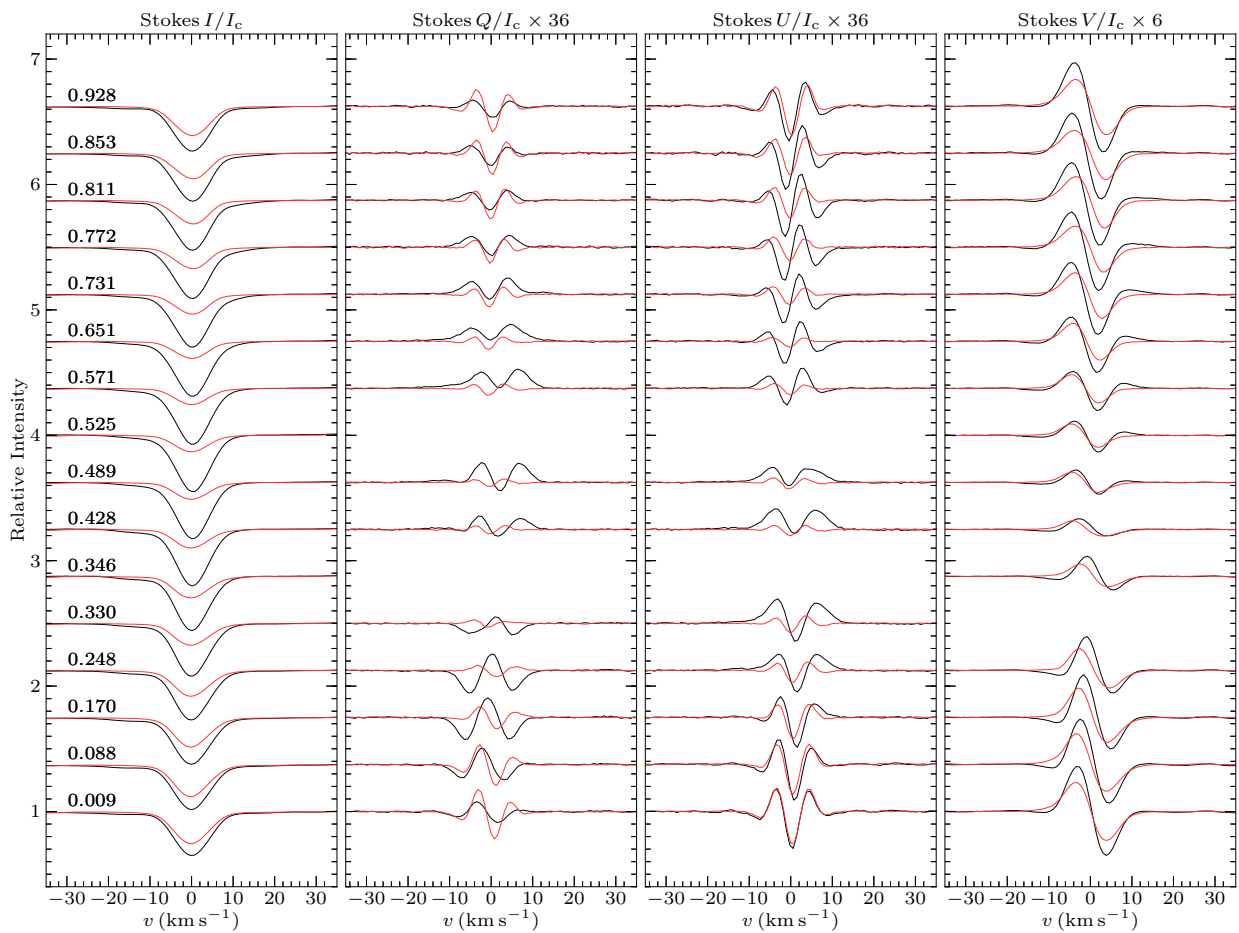}
  \caption{LSD Stokes~$I$ (\textit{first panel}), $Q$ (\textit{second panel}), $U$ (\textit{third panel}) and $V$ (\textit{fourth panel}) profiles of HD\,24712. Rotational phases calculated according to our improved ephemeris (see Sect.~\ref{sec:rot}) are indicated in the first panel. Relative to Stokes~$I$, the vertical scale for the Stokes~$Q$ and $U$ profiles has been expanded by a factor of 36. The scale for the Stokes~$V$ profiles has been expanded by a factor of 6. The error bars are on the order of the thickness of the lines, therefore they are not shown on the plots. The red lines represent the LSD profiles of REEs while the black lines show the LSD profiles of Fe-peak elements. All profiles were corrected for the mean radial velocity of the star.}
  \label{fig:LSDprofiles}
\end{figure*}

Our final line mask in combination with the multiprofile LSD code was used to obtain the LSD Stokes profiles of Fe-peak and rare-earth elements for all observations of HD\,24712. For comparison, we also produced the mean profiles using the entire mask and the standard, single-profile LSD method. The resulting LSD Stokes profiles for Fe-peak elements and REEs are presented in Fig.~\ref{fig:LSDprofiles}. These LSD spectra were computed on a velocity grid from $-80$\,\kms{} to +80\,\kms{}, with a step of 0.8\,\kms, which is similar to the mean velocity separation of the pixels in HARPS spectra. Since the scale of the LSD line weights is arbitrary \citep[see][]{Kochukhov10p5}, we have computed all LSD profiles using the following normalization coefficients: $\lambda_0 = 5000$\,\AA, $d_0 = 1$ and $\overline{g}_0 = 1$.

The calculated LSD Stokes profiles were visually compared to a subset of Stokes parameter profiles of individual Fe-peak element and REE lines which showed strong polarization signatures at minimum and maximum of the magnetic field and appeared unblended. To assess the blending we compared the observed Stokes~$I$ spectra to a synthetic spectrum. This synthetic spectrum was calculated based on the line list used for the LSD procedure and for the stellar parameters presented in Table~\ref{tab:stellar-parameters}. The program \textsc{Synth3} \citep{Kochukhov07p109} was used to produce the synthetic spectrum. 

This comparison confirms that the average polarization profiles are in good agreement with the Stokes profiles of individual spectral lines. In Fig.~\ref{fig:StokesProfilesNdFe} we have plotted variations of the Stokes~$IQUV$ profiles with phase for one Fe line and one Nd line. Their behaviour is characteristic of the variations experienced by the majority of spectral lines in HD\,24712. These two lines show strong polarization signals that are not significantly distorted by blends. By comparing the Stokes profiles of these spectral lines (Fig.~\ref{fig:StokesProfilesNdFe}) to the appropriate LSD profiles (Fig.~\ref{fig:LSDprofiles}), we can see that the latter closely match the shape and the phase behaviour of individual spectral lines. We have not noticed a qualitatively different behaviour of the LSD profiles when compared to the Stokes profiles of individual lines with different amplitudes of the polarization signatures. Therefore, we can conclude that the LSD method reproduces very well the general morphology of the polarization signatures observed in individual spectral lines.

\subsection{Choice of weights for LSD $Q$ and $U$ profiles}
\label{ssec:lsd:alt-scaling-factor}

Previous four Stokes parameter studies of Ap stars \citep{Wade00p823,Silvester2012} adopted the square of the effective Land\'e factor, $\overline{g}^2$, as the LSD weight in reconstruction of the mean Stokes~$Q$ and $U$ profiles. This essentially implies that all lines were assumed to behave as normal triplets.  However, based on the experiments with synthetic spectra, \citet{Kochukhov10p5} have argued that the use of the linear polarization sensitivity parameter $\overline{G}$, introduced by \citet{degl2004polarization}, as an LSD weight leads to a better approximation of the linear polarization spectra. 

In order to investigate how the choice of the LSD weight can affect the mean linear polarization profiles, we compared LSD $Q$ and $U$ calculations using $\overline{g}^2$ and $\overline{G}$. This tests was based on the full line mask. We found that the LSD Stokes profiles calculated with the alternative linear polarization weight, $\overline{G}$, differ by less than 5\% in relative amplitude in comparison to the LSD Stokes profiles obtained with $\overline{g}^2$. At the same time, the $\chi^2$ of the fit of the LSD model to observations changes by less than 2\%. Therefore, we can conclude that the use of the alternative scaling factor in the reconstruction of the linear polarization LSD Stokes profiles does not affect the results in a significant way. For that reason, and in order to maintain consistency with previous studies, in the rest of our LSD $Q$ and $U$ profile calculations we used $\overline{g}^2$ as suggested by \citet{Wade00p823}.

\section{Magnetic field measurements}
\label{sec:mag}

\subsection{Longitudinal field and net linear polarization from LSD profiles}
\label{ssec:mag:LSD}

It is straightforward to derive several useful quantities from the LSD Stokes profiles and corresponding null profiles. For example, one can determine the mean longitudinal component of the magnetic field and the net linear polarization \citep{Wade00p851}. The mean longitudinal magnetic field, \bz{}, represents the component of the magnetic field directed at the observer, weighted by local line strength and continuum intensity and integrated across the visible stellar disk. It can be estimated from the first-order moment of the Stokes~$V$ profile \citep{Kochukhov10p5}:
\begin{equation}
  \langle B_\mathrm{z} \rangle = -7.145 \times 10^6 \frac{\int v V(v)\,dv}{\lambda_0 \overline{g}_0 \int [1 - I(v)]\,dv},
  \label{eq:bz}
\end{equation}
where \bz{} is measured in Gauss, $\lambda_0$ in \AA{}; $V(v)$ and $I(v)$ are the LSD Stokes~$V$ and $I$ profiles. The parameters $\lambda_0$ and $\overline{g}_0$ are the same as those used for the normalization of LSD weights.

The net linear polarization is calculated from the equivalent width of the LSD Stokes~$Q$, $U$ and $I$ profiles:
\begin{equation}
  P_Q = \frac{\langle w_{QU} \rangle \int Q(v)\,dv}{\int [1 - I(v)]\,dv} \mbox{ and } P_U = \frac{\langle w_{QU} \rangle \int U(v)\,dv}{\int [1 - I(v)]\,dv}.
  \label{eq:pqu}
\end{equation}
Here $\langle w_{QU} \rangle$ represent the mean Stokes~$QU$ LSD weights. Net linear polarization contains similar information on the transverse component of the stellar magnetic field as the broad-band linear polarization \citep{Leroy1995p79}. However, mechanisms responsible for the net line and broad-band linear polarization in Ap stars are not identical, so in practice LSD $QU$ measurements need to be scaled and shifted for detailed comparison with photopolarimetric measurements \citep[e.g.,][]{Wade00p851,Silvester2012}. 

Integrals in Eqs.~(\ref{eq:bz}) and (\ref{eq:pqu}) were evaluated numerically within $\pm18$\,\kms{} limits of the mean radial velocity $21.76 \pm 0.01$\,\kms. The mean radial velocity was estimated from the Stokes~$I$ LSD profiles reconstructed with the full mask. The integration limits were chosen in a way so that the integration window includes full LSD polarization profiles. We used Eq.~(\ref{eq:bz}) to calculate \bz{} from the LSD Stokes profiles of Fe-peak elements, REEs and from the LSD profiles obtained with the full mask. For measuring net linear polarization from the LSD $Q$ and $U$ profiles we used Eq.~(\ref{eq:pqu}). All our magnetic field measurements are presented in Table~\ref{tab:bz-LSD}. Measurement uncertainties were estimated from the LSD profile variance using standard error propagation principles. For the longitudinal field measurements we achieve a typical uncertainty below 10~G for the LSD profiles produced with full and Fe-peak masks and 10--14~G for the LSD profiles of REEs.

\begin{table*}[!t]
  \caption{Magnetic field measurements obtained from the LSD Stokes profiles of HD\,24712.}
  \centering
  {\small
  \begin{tabular}{ccrrrrrrrrr} 
    \hline
    \hline
    && \multicolumn{3}{c}{$\langle B_\mathrm{z} \rangle \pm \sigma_{\langle B_\mathrm{z} \rangle}\,(\mathrm{G})$} &\multicolumn{3}{c}{$P_Q \pm \sigma_{P_Q}\,(\times10^{-5})$} & \multicolumn{3}{c}{$P_U \pm \sigma_{P_U}\,(\times10^{-5})$}\\
    HJD\,(2\,455\,000+) & Phase & \multicolumn{1}{c}{full} &  \multicolumn{1}{c}{Fe-peak} &  \multicolumn{1}{c}{REE} &  \multicolumn{1}{c}{full} &  \multicolumn{1}{c}{Fe-peak} &  \multicolumn{1}{c}{REE} &  \multicolumn{1}{c}{full} &  \multicolumn{1}{c}{Fe-peak} &  \multicolumn{1}{c}{REE} \\
    \hline
    200.71890 & 0.772 & $ 742 \pm 6$ & $465 \pm 7$ & $1304 \pm 14$ & $ 247 \pm 9$ & $ 413 \pm 10 $ & $-161 \pm  22$ & $-322 \pm 9$ & $-486 \pm  9$ & $ 104 \pm  20$\\
    201.73519 & 0.853 & $ 946 \pm 7$ & $728 \pm 8$ & $1299 \pm 13$ & $ -46 \pm 9$ & $ -10 \pm 10 $ & $-120 \pm  18$ & $-210 \pm 9$ & $-340 \pm 11$ & $  38 \pm  19$\\
    202.66063 & 0.928 & $1064 \pm 7$ & $947 \pm 9$ & $1237 \pm 11$ & $-163 \pm 7$ & $-108 \pm  9 $ & $-249 \pm  12$ & $ -92 \pm 9$ & $-182 \pm 12$ & $  33 \pm  16$\\
    203.66990 & 0.009 & $1059 \pm 6$ & $957 \pm 9$ & $1217 \pm 10$ & $-210 \pm 7$ & $-175 \pm  9 $ & $-272 \pm  11$ & $ -17 \pm 8$ & $-115 \pm 11$ & $ 114 \pm  14$\\
    204.65554 & 0.088 & $ 912 \pm 6$ & $708 \pm 8$ & $1181 \pm 10$ & $-255 \pm 6$ & $-352 \pm  8 $ & $-131 \pm  10$ & $  71 \pm 7$ & $ -21 \pm  9$ & $ 207 \pm  12$\\
    205.68041 & 0.170 & $ 720 \pm 6$ & $460 \pm 8$ & $1080 \pm 10$ & $-353 \pm 7$ & $-637 \pm  8 $ & $ 110 \pm  12$ & $ 119 \pm 7$ & $ 115 \pm  9$ & $ 127 \pm  13$\\
    206.65612 & 0.248 & $ 512 \pm 5$ & $287 \pm 6$ & $ 865 \pm  9$ & $-454 \pm 6$ & $-743 \pm  7 $ & $  73 \pm  12$ & $ 317 \pm 6$ & $ 487 \pm  7$ & $  10 \pm  11$\\
    207.67805 & 0.330 &              &             &               & $-252 \pm 6$ & $-413 \pm  7 $ & $ 110 \pm  13$ & $ 669 \pm 7$ & $ 958 \pm  8$ & $  24 \pm  16$\\
    209.65219 & 0.489 & $ 219 \pm 3$ & $ 55 \pm 3$ & $ 637 \pm  8$ & $ 556 \pm 6$ & $ 741 \pm  6 $ & $  -8 \pm  17$ & $ 564 \pm 6$ & $ 760 \pm  6$ & $ -53 \pm  17$\\
    210.66999 & 0.570 & $ 305 \pm 4$ & $ 92 \pm 4$ & $ 873 \pm 11$ & $ 704 \pm 7$ & $ 965 \pm  6 $ & $-171 \pm  18$ & $ 199 \pm 6$ & $ 265 \pm  6$ & $ -18 \pm  18$\\
    211.66838 & 0.651 & $ 465 \pm 5$ & $210 \pm 5$ & $1124 \pm 13$ & $ 636 \pm 7$ & $ 894 \pm  6 $ & $-188 \pm  18$ & $ -96 \pm 6$ & $-122 \pm  6$ & $ -16 \pm  17$\\
    212.66677 & 0.731 & $ 640 \pm 6$ & $363 \pm 6$ & $1255 \pm 14$ & $ 396 \pm 6$ & $ 627 \pm  6 $ & $-231 \pm  15$ & $-325 \pm 6$ & $-462 \pm  6$ & $  40 \pm  15$\\
    213.67090 & 0.811 & $ 837 \pm 6$ & $578 \pm 7$ & $1302 \pm 13$ & $  77 \pm 6$ & $ 177 \pm  6 $ & $-141 \pm  12$ & $-262 \pm 7$ & $-432 \pm  8$ & $ 113 \pm  16$\\
    421.89575 & 0.525 & $ 251 \pm 3$ & $ 67 \pm 4$ & $ 751 \pm 11$ &              &                &                &              &               &               \\
    606.53495 & 0.346 & $ 321 \pm 4$ & $161 \pm 4$ & $ 635 \pm  9$ &              &                &                &              &               &               \\
    607.55411 & 0.428 & $ 223 \pm 3$ & $ 76 \pm 3$ & $ 557 \pm  9$ & $ 280 \pm 7$ & $ 395 \pm  7 $ & $ -28 \pm  18$ & $ 752 \pm 9$ & $1050 \pm  9$ & $-106 \pm  22$\\
    \hline
  \end{tabular}
  }
  \label{tab:bz-LSD}
  \tablefoot{The first and second columns list heliocentric JD and rotational phase, calculated according with the improved period (Sect.~\ref{sec:rot}). Columns 3--5 provide \bz{} measurements for the Stokes $V$ LSD profiles obtained with the full mask, for the lines of Fe-peak elements and for the REE lines. Columns 6--8 report $P_Q$ measurements for the same three sets of LSD profiles and columns 9--11 provide corresponding $P_U$ measurements.}
\end{table*}

\begin{table}[!th]
  \caption{Mean and standard deviation of the magnetic field measurements obtained from null LSD profiles.}
  \centering
  \begin{tabular}{lrrr}
    \hline
    \hline
    quantity & \multicolumn{1}{c}{full mask} & \multicolumn{1}{c}{Fe-peak} & \multicolumn{1}{c}{REE} \\
    \hline
    \bz{}\, (G)               &$1.2 \pm 3 $ & $0.5 \pm 3 $ & $2.9 \pm 6$\\
    $P_Q$\, $(\times10^{-5})$ &$1.9 \pm 11 $ & $0.4 \pm 13 $ & $5.1 \pm 32$\\
    $P_U$\, $(\times10^{-5})$ &$-0.1 \pm 12 $ & $-1.3\pm 6 $ & $2.3 \pm 28$\\
    \hline
  \end{tabular}
  \label{tab:mean-Null-1sigma-errors}
\end{table}

In Table~\ref{tab:mean-Null-1sigma-errors} we report the mean and standard deviation of the \bz, $P_Q$ and $P_U$ measurements inferred from the corresponding null LSD profiles. It is evident that despite the presence of spurious polarization signals, the moment analysis of the LSD $VQU$ spectra is not affected. For example, the spurious longitudinal field amplitude derived from the Stokes $V$ null LSD profile is well below the error bar of \bz\ estimated from the LSD Stokes $V$ profiles themselves. Similarly, the spurious contribution to $P_Q$ and $P_U$ does not exceed the uncertainty due to finite $S/N$.

\subsection{Longitudinal field from the hydrogen line cores}
\label{ssec:mag:Hlines}

\begin{figure}
  \centering
  \includegraphics[width=0.4925\textwidth]{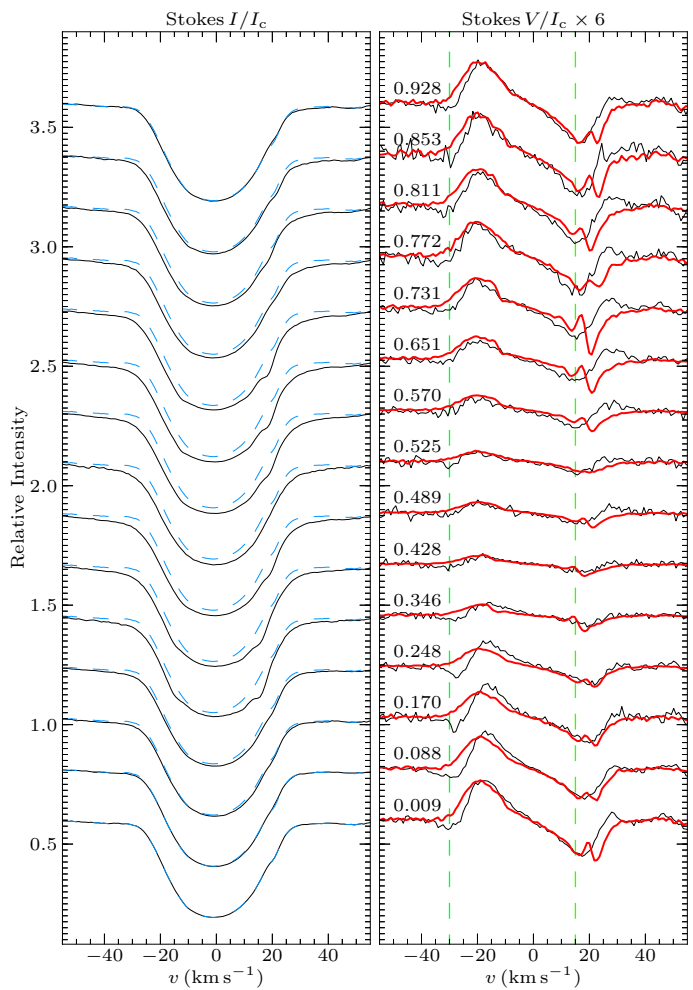}
  \caption{Stokes $I$ (\textit{left panel}) and $V$ (\textit{right panel}) spectra of the H$\alpha$ core in HD\,24712. Individual intensity spectra (\textit{solid lines}) shown in the left panel are compared with the spectrum corresponding to phase 0.009 (\textit{dashed line}). The right panel compares the observed Stokes~$V$ signal (\textit{thin line}) with the weak field model fit (\textit{thick line}). The spectral detail affecting the intensity profiles at $v\approx20$~\kms\ is due to a telluric line. Vertical dashed lines on the right panel show the velocity range used for the longitudinal magnetic field measurements.}
  \label{fig:Ha-Hb-IV-Vtheor}
\end{figure}

A high quality of the observed Stokes $V$ spectra of HD\,24712 allows us to measure the longitudinal magnetic field from the cores of hydrogen lines. This can be accomplished using two alternative methods: the moments technique, discussed in Sect.~\ref{ssec:mag:LSD} and represented by Eq.~(\ref{eq:bz}), and a differential method valid in the weak field limit: 
\begin{equation}
\label{eq:weak-field}
V(\lambda) = - k_0 \overline{g} \lambda_0^2 \langle B_\mathrm{z} \rangle \frac{dI(\lambda)}{d\lambda},
\end{equation}
where $k_0 = 4.67\,10^{-13}\,\mathrm{G}^{-1}\,\mathrm{\AA}^{-1}$, $\lambda_0$ and $\overline{g}$ are the central wavelength and effective Land\'e factor. In the case of hydrogen Balmer lines $\overline{g}=1$ \citep{Casini:1994}. To derive \bz{} with Eq.~(\ref{eq:weak-field}), we applied a weighted linear least-squares fit to Stokes $V$ spectra. 

It should be noted that, although this method of deriving \bz\ from the hydrogen lines relies on the same basic assumptions as the widely used photopolarimetric technique of \citet{Angel1970}, it is not equivalent to the latter method because it is applied to narrow cores of the hydrogen lines instead of the Stark-broadened wings. At the same time, our differential technique is more similar to the hydrogen line measurement approach used by \citet{Kudryavtsev2012p41} and \citet{Monin2012}.

We found that, when measuring \bz{} from the hydrogen line cores, the moment method can give strongly biased results due to several reasons: uncertain normalization of the Stokes~$I$ spectra around Balmer lines, difficulties in determining the limits of line cores, blending with metal lines and a non-negligible contribution from the extended wings of the Stokes~$V$ hydrogen line profiles. Regarding the latter, our theoretical calculations with the {\sc Synmast} code \citep{Kochukhov10p5} showed that the wings of the Stokes~$V$ profiles do not decrease to zero outside the core and, consequently, to get correct values for the longitudinal magnetic field one would need to integrate across the entire profile of a hydrogen line. This complicates practical application of the moment method to hydrogen lines. On the other hand, the differential method does not suffer from these limitations and, therefore, it is better suited for measuring \bz{} from the Balmer line cores in HD\,24712.

\begin{table}[!ht]
  \caption{Mean longitudinal magnetic field determined from the cores of H$\alpha$ and H$\beta$ line.}
  \centering
  \begin{tabular}{cccc}
    \hline
    \hline
    HJD\,(2\,455\,000+) & Phase & \multicolumn{2}{c}{$\langle B_\mathrm{z} \rangle$ (G)} \\
  & & H$\alpha$ & H$\beta$ \\
    \hline
    203.66990 & 0.009 & $1178 \pm 50$ & $1136 \pm 37$\\
    204.65554 & 0.088 & $ 977 \pm 61$ & $1001 \pm 28$\\
    205.68041 & 0.170 & $ 790 \pm 67$ & $ 884 \pm 36$\\
    206.65612 & 0.248 & $ 566 \pm 60$ & $ 585 \pm 32$\\
    606.53495 & 0.346 & $ 306 \pm 38$ & $ 367 \pm 22$\\
    607.55411 & 0.428 & $ 265 \pm 21$ & $ 192 \pm 29$\\
    209.65219 & 0.489 & $ 350 \pm 20$ & $ 178 \pm 43$\\
    421.89575 & 0.525 & $ 339 \pm 19$ & $ 202 \pm 52$\\
    210.66999 & 0.570 & $ 494 \pm 24$ & $ 332 \pm 49$\\
    211.66838 & 0.651 & $ 719 \pm 28$ & $ 505 \pm 80$\\
    212.66677 & 0.731 & $ 949 \pm 37$ & $ 645 \pm 76$\\
    200.71890 & 0.772 & $1116 \pm 38$ & $ 844 \pm 80$\\
    213.67090 & 0.811 & $1164 \pm 37$ & $ 953 \pm 78$\\
    201.73519 & 0.853 & $1263 \pm 55$ & $1094 \pm 95$\\
    202.66063 & 0.928 & $1284 \pm 41$ & $1098 \pm 58$\\    
    \hline
  \end{tabular}
  \label{tab:bz-Hcores}
\end{table}

The application of differential \bz\ measurement technique to H$\alpha$ is illustrated in Fig.~\ref{fig:Ha-Hb-IV-Vtheor}. This figure also demonstrates rotational modulation of the Stokes $I$ profile of this line. The consequences of this finding will be discussed below (Sect.~\ref{ssec:cwa}). The fit to the observed Stokes $V$ profiles was performed in the regions with detectable circular polarization signal. Additionally, from these wavelength regions we have excluded parts of the line cores that were affected by blends with other spectral lines. For example, for H$\beta$ the fitting limits were modified to exclude the region of a strong blend that appears on the blue side of the line core. The same was done for H$\alpha$ to exclude a telluric line on the red side of the core. The same fitting procedure and wavelength limits were also applied to the null profiles to assess systematic errors. The mean and standard deviation of the \bz{} measurements from the null profiles of H$\alpha$ equals $2 \pm 18$\,G, which indicates that possible spurious contribution to \bz\ is negligible. For H$\beta$ this value is $3 \pm 48$\,G. Results of the longitudinal field measurements from the cores of H$\alpha$ and H$\beta$ are presented in Table~\ref{tab:bz-Hcores}.

\section{Rotational period}
\label{sec:rot}
\begin{figure*}[!t]
  \centering
  \includegraphics[scale=0.80]{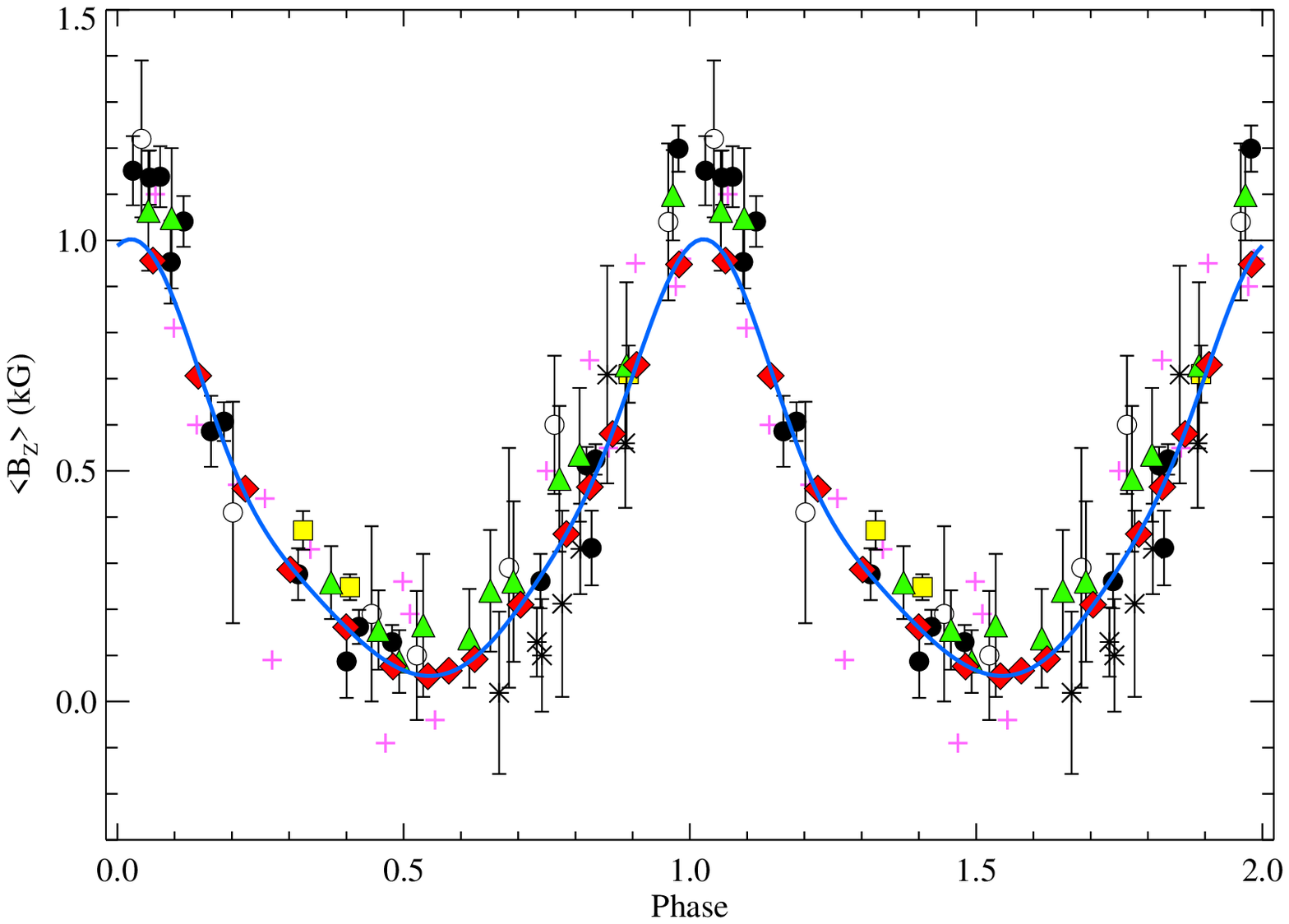}
  \caption{Variations of the mean longitudinal magnetic field of HD\,24712 phased according to the revised rotational period. Different symbols correspond to the following data sets: HARPSpol (\textit{diamonds}), \citet{Ryabchikova97p1137} (\textit{open circles}), \citet{Mathys1994p547} and \citet{Mathys97p475} (\textit{asterisks}), \citet{Leone2004p271} (\textit{squares}), measurements from TiCrFe lines by \citet{Preston1972p465} (\textit{pluses}), \citet{Wade00p851} and \citet{Ryabchikova2005p55} (\textit{filled circles}), \citet{Ryabchikova2007p1103} (\textit{triangles}). The solid line represents a 4th-order Fourier fit to the longitudinal field measurements.}
  \label{fig:RotPeriod}
\end{figure*}

The most recent determination of the rotational period of HD\,24712 was done by \citet{Ryabchikova2005p55}. They combined their Fe-Cr line longitudinal magnetic field measurements with other data from the literature, finding $P_\mathrm{rot} = 12.45902 \pm 0.00044\,\mathrm{d}$. They also obtained an independent period estimate, $P_\mathrm{rot} = 12.45853 \pm 0.00018\,\mathrm{d}$, from the variability of equivalent widths of REE lines. The final period recommended by \citet{Ryabchikova2005p55}, $P_\mathrm{rot} = 12.45877 \pm 0.00016$, represents a formal weighted mean of the periods found using the magnetic and equivalent width data. A significantly smaller period, $P_\mathrm{rot} = 12.4572 \pm 0.0003\,\mathrm{d}$, was inferred from the photometric pulsational analysis of HD\,24712 \citep{Kurtz1987p285,Kurtz2005}.


We revised rotational period of HD\,24712 taking advantage of the new longitudinal field measurements obtained here and by \citet{Ryabchikova2007p1103}. For the period search we compiled a list of all published longitudinal magnetic field measurements derived from the spectral lines of Fe-peak elements, which should be among the least variable lines in the spectrum of HD\,24712. This list includes the TiCrFe \bz\ data published by \citet{Preston1972p465}, measurements from the \ion{Fe}{i} lines by \citet{Mathys1991p121} and \citet{Mathys97p475}, longitudinal field derived from the LSD Stokes $V$ profiles of Fe-peak elements by \citet{Wade00p851} and \citet{Ryabchikova2005p55}, measurements by \citet{Leone2004p271} and, finally, measurements from the recent spectroscopic study of pulsations in the atmosphere of HD\,24712 performed by \citet{Ryabchikova2007p1103}. Photopolarimetric longitudinal magnetic field measurements from the H$\beta$ line by \citet{Ryabchikova97p1137} were also included. However, these measurements show systematically higher values of \bz{} than those derived from metallic lines. Therefore, these measurements were reduced by 400\,G following the analysis by \citet{Ryabchikova2005p55}. A similar correction of $-300$~G was applied to the data from \citet{Preston1972p465}. Error bars are lacking for this data set in the original publication. We adopted an error bar of 150~G, which is close to the standard deviation of this data set with respect to final fit. The literature data were finally complemented with our own longitudinal magnetic field measurements obtained from the Fe-peak lines (see Table~\ref{tab:bz-LSD}). The final data set used for period search comprised 81 data points, spanning the time interval from 1969 to 2011.

The period search was performed by fitting a fourth-order harmonic curve to the entire collection of the longitudinal field data. This high-order fit was required to match our measurements within their error bars. As a starting value, we used the period derived by \citet{Ryabchikova2005p55} and adopted the zero point from their paper. Our final ephemeris is given by:
$$
\mathrm{HJD}(\langle B_\mathrm{z} \rangle \, _\mathrm{max}) = 2\,453\,235.18(40) + 12.45812(19)\cdot E.
$$

This is almost the same precision as claimed by \citet{Ryabchikova2005p55} for their average period. However, here we have achieved this result by using the magnetic field data alone. Our new period agrees within the error bars the one found by \citet{Ryabchikova2005p55} from the equivalent width measurements and provides a better fit to the longitudinal field data than their average period.

\section{Global magnetic field parameters}
\label{gf-par}

\begin{figure*}[!t]
\centering
\includegraphics[width=17cm]{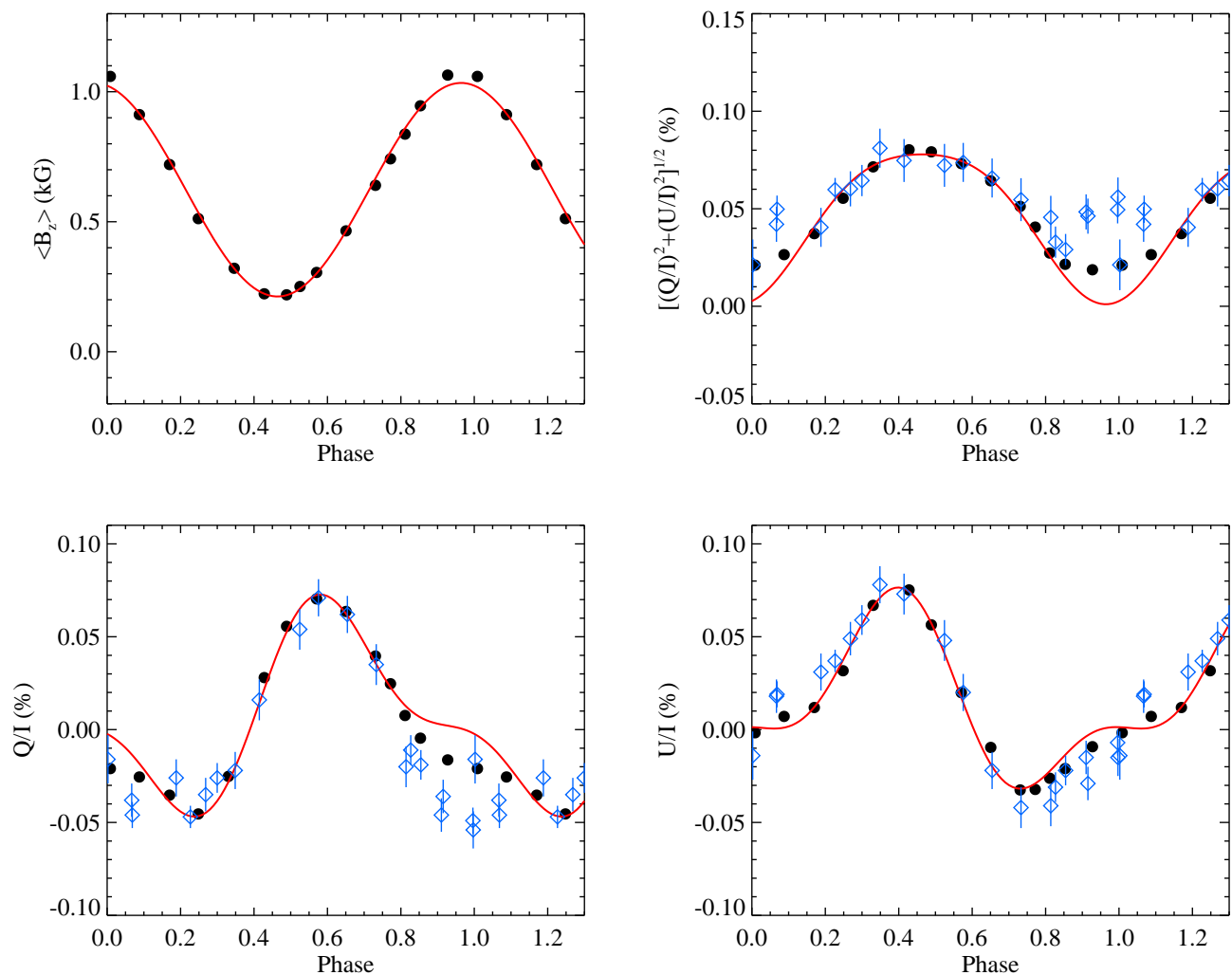}
\caption{Comparison of the canonical model predictions (\textit{thick lines}) with the magnetic measurements obtained from LSD profiles of HD\,24712 (\textit{circles}) and broadband linear polarization measurements by \citet{Leroy1995p79} (\textit{rombs}). The panels show longitudinal magnetic field (\textit{upper left}), total linear polarization (\textit{upper right}) and the phase curves of the linear polarization in the Stokes $Q$ (\textit{lower left}) and $U$ (\textit{lower right}) parameters.}
\label{fig:gf-par}
\end{figure*}

An important step in the understanding of the structure of magnetic fields of Ap stars was made with the introduction of the so-called ``canonical'' model \citep{Landolfi1993p285}. By considering  circular and linear polarization observables together, the canonical model allowed to extract a more complete information about the magnetic field geometry compared to a simple fitting of the longitudinal field curve. Indeed, the circular polarization is directly connected only to the line-of-sight magnetic field component. It is normally characterized with the mean longitudinal magnetic field, which, for a global axisymmetric magnetic field geometry, has a simple harmonic dependence on the rotational phase with the amplitude proportional to the polar field strength. On the other hand, the broadband linear polarization is sensitive to the transverse field component and its phase curve is strongly dependent on the stellar magnetic field geometry.

\citet{Landolfi1993p285} introduced the canonical model under the following assumptions: weak-field approximation, homogeneous distribution of chemical elements across the surface of the star, all lines contributing to the spectral interval are characterized by the same average strength, Doppler width, damping constant and Land\'e factor. It is under these assumptions that a derivation of analytical formulas for the phase variation of the broadband linear polarization becomes possible.

For a purely dipolar magnetic field geometry, the canonical model is simplified even further. One describes the field geometry with three angles: the angle between the rotational axis and the line of sight $i$, the angle between the magnetic axis and the rotational axis $\beta$, and the azimuth angle of the rotational axis $\Theta$. In addition to these three angles, the canonical model has three additional parameters: the polar magnetic field strength $B_{\rm p}$, the constant $\kappa$ giving the amplitude of linear polarization, and parameter $\epsilon$, which describes the magneto-optical effects. In some specific cases one might introduce three more parameters: $q_0$ and $u_0$ that are responsible for the interstellar linear polarization in the Stokes $Q$ and $U$ parameters respectively, and the angle $f_0$ indicating the phase when the dipolar axis is located in the plane defined by the stellar rotational axis and the line of sight. All in all, we have nine different parameters out which six are primary and the rest are of secondary importance.

Using magnetic field measurements inferred from the LSD profiles of HD\,24712, we derived the canonical model parameters in the same way as it was previously done by \citet{Bagnulo1995p459}. A $\chi^2$ function was constructed that connects the observed and theoretical values of the longitudinal field and net linear polarization. This $\chi^2$ function was minimized in order to find the model parameters that provide the best fit to observations. Because both the Fe-peak and rare-earth elements have different non-uniform distributions across the surface of the star, we used measurements obtained using the full line mask. Furthermore, the net linear polarization measurements computed from the Stokes $Q$ and $U$ LSD profiles were rescaled to match the amplitude of the broadband linear polarization measurements by \citet{Leroy1995p79} following the usual procedure \citep{Wade00p851,Silvester2012}.

The minimization of the $\chi^2$ function provided the following results: $i = 138\pm1\degr$, $\beta = 144 \pm 1\degr$, $\Theta = 29 \pm 3\degr$, $B_{\rm p} = 3355 \pm 31\,\mathrm{G}$, $\kappa = 0.081 \pm 0.003$, $\epsilon = 0.03 \pm 0.22$ and $f_0 = 0.036 \pm 0.002$. In this fit $q_0$ and $u_0$ were held equal to zero due to the absence of interstellar linear polarization effects in spectropolarimetric observations. Repeating parameter optimization with fixed values of $\epsilon$ had no significant effect on the model parameters except for $\Theta$, which varied by 2--3~$\sigma$ when $\epsilon$ changed in the range between 0 and 1. In Fig.~\ref{fig:gf-par} we illustrate the theoretical phase curves of the circular and linear polarization magnetic observables, together with the broadband linear polarization measurements by \citet{Leroy1995p79} and our LSD profile measurements.

The final value of the reduced $\chi^2$ is approximately 3.3. As one can see from Fig.~\ref{fig:gf-par}, the model curves noticeably deviate from the Stokes $Q$ observations in the 0.8--0.1 phase interval. The discrepancy between the broadband linear polarization measurements and the net linear polarization inferred from the LSD profiles is also maximal in this interval. It is possible that these discrepancies are due to an inhomogenous distribution of chemical elements, in particular REEs, which reach their maximum strength in this part of the phase curve.

The comparison of our results with the dipolar model parameters obtained by \citet{Bagnulo1995p459} shows a good agreement, except for a somewhat lower $B_{\rm p}$ obtained here and a difference in the azimuth angle. In our case $\Theta$ is close to 30\degr\ while Bagnulo et al. found the value around $5\degr$. This discrepancy could be traced to different rotation periods adopted in our study and by \citet{Bagnulo1995p459}. Using their (outdated) value of $P_{\rm rot}$ we could also reproduce their azimuth angle.

\section{Summary and discussion}
\label{sec:sum}

We have obtained spectra in all four Stokes parameters for the cool pulsating Ap star HD\,24712. These data, secured with the HARPSpol instrument at the ESO 3.6-m telescope, have resolving power exceeding 10$^5$ and a S/N ratio of 300--600. In total, we have obtained 43 individual Stokes parameter observations covering the entire rotational period of the star. The spectropolarimetric data set collected for HD\,24712 is the first ever full Stokes vector observation of a magnetic star at such a high quality.

The HARPSpol spectra of HD\,24712 were employed to study the morphology of the Stokes profile shapes in individual spectral lines and to calculate the LSD Stokes profiles using different line masks. In addition, we analyzed null LSD profiles and assessed spurious linear polarization signatures appearing due to the crosstalk from circular polarization. Our analysis showed that this crosstalk does not exceed 0.5\% and that the spurious polarimetric signatures are 15--60 times smaller than the corresponding LSD Stokes profiles and, therefore, do not affect interpretation of the data.

Using the LSD Stokes~$V$ profiles, we measured the mean longitudinal magnetic field with an accuracy of 5--10\,G. We also derived the net linear polarization from the LSD Stokes~$Q$ and $U$ profiles. Combining our \bz\ measurements with previous measurements taken from the literature, we improved rotational period of HD\,24712, finding the value of $P_{\mathrm{rot}} = 12.45812\pm0.00019\,\mathrm{d}$.

The new \bz\ and net linear polarization measurements were also used to determine parameters of the dipolar magnetic field topology. We found that magnetic observables can be reasonably well reproduced by the dipolar model, although noticeable discrepancies remain at some phases, which we tentatively attribute to the effects of chemical abundance spots. The high S/N ratio of the Stokes~$V$ spectra allowed us to measure the longitudinal magnetic field from the cores of H$\alpha$ and H$\beta$ lines. 

The analysis presented in this paper represents the first step towards obtaining detailed maps of the magnetic field and horizontal and vertical abundance structures for HD\,24712. 

\subsection{Vertical magnetic field gradient}
\label{ssec:disc}

\begin{figure*}[!t]
  \centering
  \includegraphics[width=0.4975\textwidth]{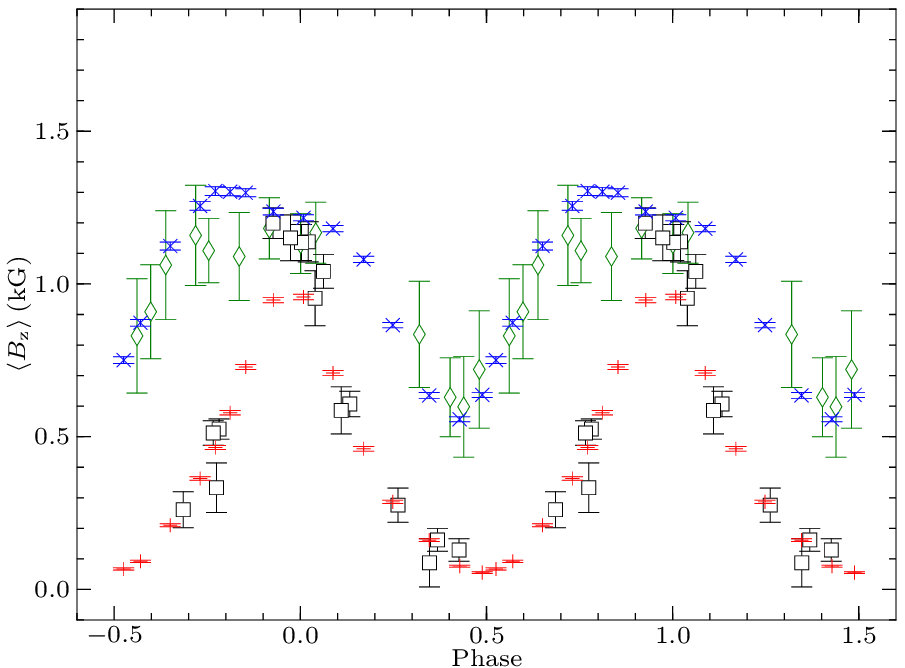}
  \includegraphics[width=0.4975\textwidth]{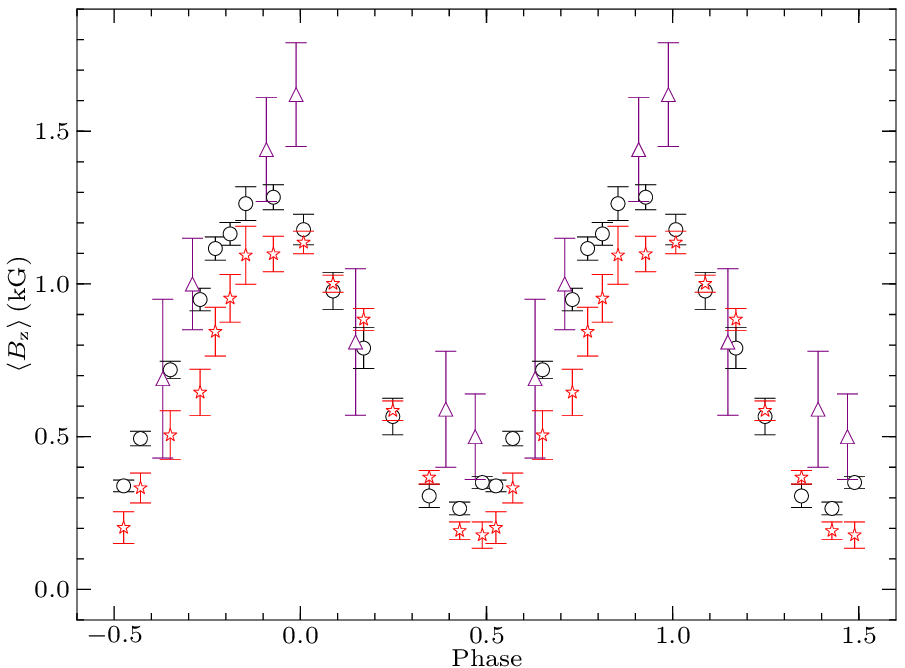}
  \caption{Mean longitudinal magnetic field measurements from the LSD Stokes profiles (\textit{left panel}) and the cores of hydrogen lines (\textit{right panel}). On the left panel the red crosses and blue ``x'' symbols correspond to our measurements for the Fe-peak and rare-earth elements. The diamonds and squares are previous measurements from the lines of REEs \citep{Ryabchikova2007p1103} and Fe-peak elements \citep{Ryabchikova2005p55} respectively. On the right panel we show the longitudinal field measurements obtained from the cores of H$\alpha$ (\textit{circles}) and H$\beta$ (\textit{stars}), and previous longitudinal magnetic field measurements from the wings of H$\beta$ by \citet{Ryabchikova97p1137} (\textit{triangles}).}
  \label{fig:bz-LSD-Hcores}
\end{figure*}

The HARPSpol spectra of HD\,24712 allow us to investigate an interesting problem of the vertical gradient of magnetic field. In the past, searches for the vertical variation of magnetic field in Ap stars were focused on measuring the longitudinal magnetic field or the mean field modulus using the spectral lines located before and after the Balmer jump. If a radial field gradient is present, lines formed at different depths in the atmosphere will show discrepant magnetic field and a different \bz{} as a function of rotational phase. A handful of investigations that used this technique \citep{Preston1967p3,Wolff1978p412,Romanyuk1986p25,Nesvacil2004p51,Romanyuk2007p26} showed no conclusive evidence for the presence of a radial field gradient in Ap stars. 

Recently, the search for vertical magnetic field gradient was extended to the analysis of hydrogen line cores. These features are formed significantly higher in comparison to metal lines, they are sufficiently wide and show easily measurable Stokes~$V$ signatures. \citet{Kudryavtsev2012p41} derived magnetic field from the H$\beta$ core for a sample of Ap stars. Their measurements showed that for the majority of stars the mean longitudinal magnetic field determined from H$\beta$ is significantly lower than \bz\ obtained from the metal lines in the same spectra. \citet{Kudryavtsev2012p41} attributed this to the existence of a radial gradient of the magnetic field. This implies that magnetic field strength in the atmospheres of Ap stars decreases in the radial direction significantly faster than expected for a simple dipolar field topology.

An exceptionally high quality of the spectropolarimetric data set acquired for HD\,24712 allowed us to derive precise \bz\ measurements for different groups of lines and to test the vertical field gradient hypothesis for this particular object. Moreover, recent magnetic Doppler imaging, pulsation tomography and model atmosphere studies of this star \citep{Luftinger10p71,Ryabchikova2007p1103,Shulyak09p879} provide additional useful constraints on the horizontal and vertical formation regions of the spectral lines belonging to different chemical species.

If a radial magnetic field gradient is present, a difference between magnetic field measurements from the Fe-peak elements, REEs and the cores of hydrogen lines might be expected. However, Ap stars in general and HD\,24712 in particular possess strongly inhomogeneous horizontal distributions of chemical elements in their atmospheres. Hence, if we want to detect a radial field gradient by comparing magnetic field inferred from different groups of metallic lines, we would first need to remove the effects of spots, which significantly complicates analysis of \bz\ measurements.

Balmer lines, on the other hand, enable a more robust radial field gradient detection. The formation height of the cores of Balmer lines decreases from H$\alpha$ to higher members of the series. In the case of HD\,24712 this is confirmed by the pulsation amplitudes increase from H$\gamma$ to H$\alpha$ \citep{Ryabchikova2007p1103}. This means that the magnetic field measured from, e.g., H$\beta$ should be stronger than the field measured from H$\alpha$ if magnetic field \textit{decreases} with height.

Our magnetic field measurements from the LSD profiles and from the cores of hydrogen lines are presented in Fig.~\ref{fig:bz-LSD-Hcores}. This figure also compares our data with previous measurements by other authors. A visual comparison between the magnetic phase curves obtained from the Fe-peak and REEs lines reveals a large difference. Taking into account that REE absorption features form at significantly higher atmospheric layers than Fe-peak lines \citep{Ryabchikova2007p1103,Shulyak09p879}, one might naively conclude that the field strength \textit{increases} with height.

Instead, it is more likely that this difference is entirely due to an inhomogeneous horizontal distribution of chemical elements. \citet{Luftinger10p71} demonstrated that HD\,24712 exhibits an anti-correlation between the local abundance enhancement of the rare-earth and Fe-peak elements, with the maximum local abundance of REEs and minimum abundance of Fe-peak elements roughly coinciding with the positive magnetic pole. This abundance distribution should exactly lead to the picture that we see in the left panel of Fig.~\ref{fig:bz-LSD-Hcores}.

Magnetic field measurements from the cores of hydrogen lines lie in between the REE and Fe-peak \bz\ curves and are more sinusoidal than the latter -- just as expected from a much weaker rotational modulation of the hydrogen lines. However, the \bz{} curves derived from the cores of H$\alpha$ and H$\beta$ lines do show small discrepancies between them. For phases less than 0.5 we see that both \bz\ curves are identical to within error bars, but for phases greater than 0.5, the \bz\ curve for H$\alpha$ lies slightly higher than the one for H$\beta$. 
One should also note that in this phase interval the \bz{} measurements for H$\beta$ have a greater standard deviation since the derivative of Stokes $I$ cannot exactly fit the shape of Stokes $V$.
Here we should also mention that previous photopolarimetric longitudinal magnetic field measurements obtained from the wings of H$\beta$ by \citet{Ryabchikova97p1137} are systematically stronger than our line-core measurements by about 200\,G. This is likely due to an uncertainty of the calibration and interpretation of the hydrogen-line photopolarimetric line-wing measurements \citep{Mathys2000}.

We conclude that our data do not present an unambiguous evidence for a radial magnetic field gradient in the atmosphere of HD\,24712. It is obvious that magnetic field measurements from considering individual lines are inherently ill suited for detecting subtle effects such as vertical magnetic field gradient. The small discrepancies present in the \bz{} measurements from the cores of hydrogen lines, instead, further strengthen our case for obtaining self-consistent detailed maps of the magnetic field and horizontal and vertical abundance structures for HD\,24712. 

\subsection{Rotational modulation of the hydrogen core-wing anomaly}
\label{ssec:cwa}

In Sect.~\ref{ssec:mag:Hlines} we have demonstrated the existence of the rotational modulation of the H$\alpha$ line core in HD\,24712. This discovery has interesting implications for understanding the interplay between diffusion and atmospheric structure of chemically peculiar stars. Many cool Ap stars exhibit the so-called core-wing anomaly \citep[CWA,][]{Cowley2001} in the hydrogen line profiles. This feature, which is most clearly seen in H$\alpha$, is manifested as an unusually sharp transition between the Stark-broadened wings and Doppler core when compared to normal stars of similar spectral types or to spectrum synthesis calculations with conventional stellar model atmospheres. The width of the Doppler core also appears to be too narrow. \citet{Kochukhov2002} were able to fit the CWA in several cool Ap stars by empirically modifying the $T$-$\tau$ relation. They showed that in order to reproduce the hydrogen line shape one needs a temperature increase by 500--1000~K between $\log\tau_{5000}=-1$ and $-
4$. A qualitatively similar modification of the atmospheric structure is obtained by the self-consistent theoretical \citep{LeBlanc2009} or empirical \citep{Shulyak09p879} diffusion models. In particular, the latter study obtained an inverse temperature gradient at $\log\tau_{5000}\approx-3$ as a consequence of a REE-rich layer located in the upper atmosphere of HD\,24712. Thus, the CWA provides a unique probe into the anomalous structure of the upper atmospheres of cool Ap stars.

Here we showed, for the first time, that the CWA exhibits rotational variability, synchronized with that of magnetic field and metal line strengths. Our data shows that the CWA is less pronounced (the core width is larger and the core-wing transition is smoother) at phase $\approx$\,0.5, corresponding to the minimum magnetic field and maximum strength of the Fe-peak element lines. Conversely, the anomaly is largest at phase 0.0, which coincides with the maximum magnetic field and maximum REE line strength. This observation strengthens the connection between the CWA and stratification of REEs proposed by \citet{Shulyak09p879}. The distribution of rare-earth elements is modulated in height and across the stellar surface in a such a way that we observe the largest atmospheric anomaly when the REE-rich layer in the upper atmosphere comes into view. 

At the same time, this picture is not fully consistent with the theoretical diffusion predictions \citep{LeBlanc2009,Alecian2010}. Current theoretical models correctly predict an accumulation of the Fe-peak elements in the horizontal field regions at the magnetic equator. However, this is accomplished in a very specific way, by an increase of the element abundance at high-atmospheric layers. As argued by \citet{LeBlanc2009}, this should heat the upper atmospheric layers and produce a CWA-like distortion in the hydrogen lines. But for HD\,24712 this scenario implies the highest upper atmosphere abundance of all elements and hence the largest atmospheric anomaly and the strongest CWA at the magnetic minimum phase, when the field on the visible surface is mostly horizontal. This is opposite to what is observed. Thus, either the current equilibrium diffusion models are unrealistic due to a missing physical ingredient or incorrect basic assumptions, or the impact of the hitherto unexplained REE stratification on the atmospheric structure completely dominates the observations.

\begin{acknowledgements}
OK is a Royal Swedish Academy of Sciences Research Fellow, supported by grants from Knut and Alice Wallenberg Foundation and Swedish Research Council.
\end{acknowledgements}

\bibliographystyle{aa}
\bibliography{articles}

\Online
\begin{appendix}
\section{Crosstalk}
\label{sec:crosstalk}

Previous HARPSpol observations demonstrated an exceptional polarimetric sensitivity of this instrument. For example, \citet{Piskunov11p7} could put an upper limit of $\sim$\,10$^{-5}$ on the circular and linear polarization signal in the lines of the bright inactive star $\alpha$\,Cen\,A. These results demonstrate that the instrument does not suffer from a noticeable spurious polarization. Intrinsically unpolarized spectral features, such as telluric lines, also show no polarization artifacts down to $\sim$\,10$^{-3}$ in our observations of HD\,24712. However, these measurements do not directly constrain the polarimetric precision, i.e. the uncertainty with which strong polarization signals can be measured with this instrument. Our observational data for HD\,24712 and the employed polarimetric demodulation method allow us to derive not only the Stokes parameters, but also the null polarization spectra. The latter can be used as a diagnostic tool to detect and characterize spurious polarization and crosstalks. In this section we present a detailed analysis of the mean Stokes $VQU$ null profiles.

\begin{figure}[!t]
  \centering
  \includegraphics[width=0.4925\textwidth]{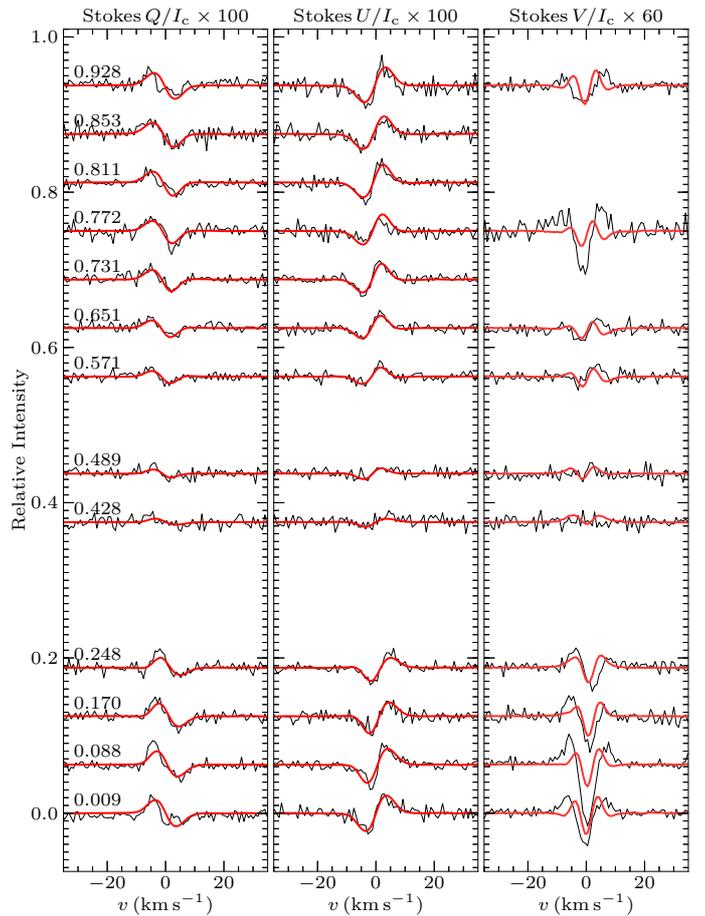}
  \caption{Analysis of the LSD null profiles of HD\,24712. Thin black lines show mean profiles derived from the null Stokes $Q$ (\textit{left panel}), $U$ (\textit{middle panel}), and $V$ (\textit{right panel}) spectra. The thick red lines show an attempt to reproduce these spurious polarization signatures with scaled LSD Stokes $V$ profiles ($QU$ panels) and a linear combination of the LSD Stokes $Q$ and $U$ profiles ($V$ panel).}
  \label{fig:crosstalk}
\end{figure}

With the help of the multiprofile LSD code we calculated LSD null profiles for Fe-peak elements, REEs and for the full line mask. The results of these calculations, illustrated in Fig.~\ref{fig:crosstalk}, demonstrate the presence of very weak signals in all of the null LSD profiles. The amplitude of these signals in the LSD~$Q$, $U$ and $V$ null profiles is 12, 25 and 60 times smaller than the corresponding LSD Stokes profiles. The signals present in the null $Q$ and $U$ profiles are similar in shape and correlate in amplitude with the Stokes $V$ signal. This might be an indication of a crosstalk from circular to linear polarization. 

On the other hand, the null LSD~$V$ profiles show a more complicated picture that most likely is a combination of several factors. Instead of a clear rotational phase dependence as in the case of the null LSD~$Q$ and $U$ profiles, we see only a semi-regular variability. Thus, we cannot attribute the null Stokes $V$ signal to a crosstalk from $Q$ and $U$ to $V$. An alternative possibility is that the Stokes $V$ null spectrum is dominated by a spurious signature appearing due to a variation of the barycentric velocity in the course of a fairly long sequence of polarimetric sub-exposures. In that case, the Stokes $V$ null signature would be proportional to the derivative of Stokes $V$ profile itself times the velocity change between the first and last sub-exposure.

We used a weighted linear least-squares fitting procedure to estimate the crosstalk levels. In the case of circular to linear polarization we found a crosstalk level of $0.37 \pm 0.01$\,\% for Stokes~$Q$ and $0.5 \pm 0.01$\,\% for Stokes~$U$. Figure~\ref{fig:crosstalk} shows that the null linear polarization signatures are indeed very well approximated by the scaled Stokes $V$ profile. A similar fit attempt to reproduce the null Stokes $V$ profile by a superposition of the Stokes $Q$ and $U$ profiles is less successful and requires crosstalk levels of 4--8\%.

Summarizing this crosstalk analysis, we can state that the detected spurious polarimetric signals can not significantly affect our measurements because their level is 15 to 60 times smaller than the corresponding LSD Stokes profiles. Nevertheless, in the subsequent measurements using the LSD Stokes profiles (Sect.~\ref{ssec:mag:LSD}) we continue using null profiles to assess possible systematic errors due to a non-ideal polarimetric measurements and artifacts in data processing. 

\end{appendix}

\end{document}